\documentclass{aa}

\usepackage{graphicx}
\usepackage{txfonts}
\usepackage{lipsum}
\usepackage{subcaption}         
\usepackage{lscape}             
\usepackage{placeins}           
                                
\usepackage{natbib,twoopt}
\usepackage[breaklinks=true]{hyperref}
\usepackage{url}
\usepackage{comment}
\usepackage{pdfpages}
\usepackage[normalem]{ulem}
\usepackage{xcolor}
\usepackage{enumitem}
\usepackage{upgreek}
\usepackage{threeparttable}

\usepackage{nolinenumbers}

\bibpunct{(}{)}{;}{a}{}{,}

\newcommand{\mum}{\textmu{}m\xspace}
\newcommand{\HI}{\textrm{H~{\textsc{i}}}\xspace}
\newcommand{\HII}{\textrm{H~{\textsc{ii}}}\xspace}
\newcommand{\molh}{H$_2$\xspace}

\newcommand{\classA}{$\mathcal{A}$\xspace}
\newcommand{\classB}{$\mathcal{B}$\xspace}
\newcommand{\classAB}{$\mathcal{AB}$\xspace}
\newcommand{\classC}{$\mathcal{C}$\xspace}
\newcommand{\classD}{$\mathcal{D}$\xspace}

\begin{document}

   \title{PAH Spectral Diversity in NGC~7027 and the Evolution of Aromatic Carriers}


   \author{Charlotte Smith-Perez\inst{\ref{WOntarioPA} } \and  
   Aidan Hembruff\inst{\ref{WOntarioPA}} \and  
   Els Peeters\inst{\ref{WOntarioPA},~\ref{WOntarioIESE}} \and
    Alexander G.~G.~M. Tielens \inst{\ref{Leiden},~\ref{Maryland}} 
    \and
    Alessandra Ricca
    \inst{\ref{Ames}, ~\ref{SETI}}}
 
   \institute{
    Department of Physics \& Astronomy, The University of Western Ontario, London ON N6A 3K7, Canada
    \label{WOntarioPA}\\ \email{epeeters@uwo.ca} \and
    Institute for Earth and Space Exploration, The University of Western Ontario, London ON N6A 3K7, Canada
    \label{WOntarioIESE}  \and 
    Leiden Observatory, Leiden University, P.O. Box 9513, 2300 RA Leiden, The Netherlands
    \label{Leiden} \and
    Astronomy Department, University of Maryland, College Park, MD 20742, USA
    \label{Maryland} \and
    NASA Ames Research Center, MS 245-6, Moffett Field, CA 94035-1000, USA 
    \label{Ames} \and
    Carl Sagan Center, SETI Institute, 339 Bernardo Avenue, Suite 200, Mountain View, CA 94043, USA
   \label{SETI}
  }
             
\titlerunning{PAH Spectral Diversity in NGC~7027}
\authorrunning{Smith-Perez et al.}             

   \date{Received October 10, 20XX}

 
  \abstract
   {Polycyclic Aromatic Hydrocarbons (PAHs) constitute a significant fraction of the Universe's carbon budget, playing a key role in the cosmic carbon cycle and dominating the mid-infrared spectra of astrophysical environments in which they reside. Although PAHs are known to form in the circumstellar envelopes of post-AGB stars, their formation and evolution are still not well-understood.}
   {We aim to understand how pristine complex hydrocarbons and PAHs in circumstellar environments transition to the PAHs observed in the ISM. } 
   {The mid-infrared PAH spectra (5-18~\mum) of the planetary nebula, NGC~7027, are investigated using spectral cubes from JWST MIRI-MRS.}
   {We report the first detection of spatially-resolved variations of the PAH spectral profiles across class \classA, \classAB, and \classB in all major PAH bands (6.2, 7.7, 8.6, and 11.2~\mum) within a single source, NGC~7027. These variations are linked to morphological structures within NGC~7027. Clear correlations are revealed between the 6.2, 7.7, and 8.6~\mum features, where the red components (6.26, 7.8, 8.65~\mum) exhibit a strong correlation and the same is found for the blue components of the 6.2 and 7.7~\mum (6.205 and 7.6~\mum). The blue component of the 8.6 (8.56~\mum) appears to be independent of the other components. We link this behaviour to differences in molecular structure of their PAH subpopulations. Decomposition of the 11.2~\mum band confirms two previously identified components, with the broader 11.25~\mum component attributed to emission from very small grains or PAH clusters rather than PAH emission.}
   {We show that PAH profile classes generally vary with proximity to the central star's UV radiation field, suggesting class \classB PAHs represent more processed species while class \classA PAHs remain relatively pristine, challenging current notions on the spectral evolution of PAHs.}


   \maketitle

\section{Introduction}
\label{sec:intro}


The outflows of dying intermediate-mass stars (i.e., AGB stars) are the birth site for dust and carbon-related species, such as PAHs \citep[e.g.,][]{Frenklach1989, Boersma2006, Galliano2008a, Tielens2008}. Carbon-rich AGB stars dictate the early chemistry in their circumstellar envelopes before the molecules and dust are later incorporated into the interstellar medium (ISM), providing the main source of carbon and other organic elements for the ISM \citep[e.g.,][]{Cherchneff1992, Tielens2005, Matsuura2009}. Therefore, PAH emission from post-AGB stars and planetary nebulae (PNe) can be studied to unveil the PAH characteristics at the beginning stages of their lifecycle. 

The characteristics of the PAH population are strongly influenced by the temperature, density, and radiation field of the environment in which they reside. It follows that the PAH emission bands reflect the conditions of their environment, and spectral variations arise in the form of relative intensity, peak emission position, and profile shape \citep[e.g.,][]{Cohen1986, Smith2007, Galliano2008b}.  

Based on observed peak positions and profile shapes of the 3.3, 6.2, 7.7, and 11.2~\mum aromatic infrared bands (AIBs), spectra can be categorized in different classes: \classA–\classD \citep{Peeters2002, vanDiedenhoven:chvscc:04, Matsuura2014,  Sloan2014}. These different AIB classes correlate with source type \citep[i.e., \HII regions, reflection nebulae, Herbig AeBe stars, planetary nebulae;][]{Cohen1989a, Bregman1989, Peeters2002, vanDiedenhoven:chvscc:04} and the observed spectral differences are thought to reflect processing of the PAHs in the local environment by prevalent FUV photons, shocks, and/or gas phase chemistry \citep[e.g.,][]{Peeters2002, Boersma2008, Matsuura2014}. Likely, there is an evolutionary relationship between these classes. Indeed, it has been suggested that recently formed PAHs are predominantly class \classB but processing in the harsh environment of the ISM converts them into class \classA PAHs \citep[e.g.,][]{Peeters2002, Boersma2008, Matsuura2014}. A somewhat different scenario is envisioned by \citet{Joblin2008} in which PAHs formed in C-rich AGB outflows are rich in aliphatics and these species are converted into highly aromatic compounds in harsh radiation fields. There is observational evidence that this conversion already occurs in the (proto)-planetary nebula phase \citep{goto2003,goto2007}.

Infrared studies with ISO-SWS have revealed that the planetary nebula, NGC~7027, has an AIB spectrum that is a mix of class \classA and \classB \citep{Peeters2002, vanDiedenhoven:chvscc:04}. This source provides therefore a unique opportunity to assess the factors that drive the conversion from class \classB to class \classA. However, the beam size of the SWS instrument was similar to the size of this source in the mid-IR and thus precluded a detailed study of the spatial distribution of the AIBs to trace this evolution in the outflow. The MIRI IFU \citep{MIRI, MRS} on board JWST \citep{jwst} with a pixel scale of 0.2" is well-suited to study the spatial-spectral distribution of the AIBs in this nebula.

In Section~\ref{sec:AIB_class}, we describe the AIB classification and in Section~\ref{sec:7027}, we discuss the morphology of NGC~7027. Details about the observations, data reduction, and analysis are discussed in Section~\ref{sec:obs_data_reduction}. We present the results in Section~\ref{sec:results}, followed by a discussion in Section~\ref{sec:discussion}. Finally, the conclusions of the study are laid out in Section~\ref{sec:conclusions}. 

\section{AIB Classification}
\label{sec:AIB_class}

Class \classA profiles exhibit the bluest peak positions. For class \classA profiles, the 6.2~\mum band has a peak position between 6.19 and 6.23~\mum, the 7.6~\mum feature is the dominant component of the 7.7~\mum complex, and the 8.6~\mum feature peaks between 8.58 and 8.62~\mum \citep{Peeters2002}. For the 11.2~\mum band, a class \classA or \classAB profile peaks at 11.20-11.24~\mum. The difference between the 11.2~\mum \classA and \classAB classifications arise from the full width at half maximum (FWHM), where a class \classA profile has a narrower profile than that of class \classAB \citep{vanDiedenhoven:chvscc:04}. Class \classB, on the other hand, demonstrates red-shifted peak positions. In this case, the 7.8 is the dominant component of the 7.7~\mum complex and the 8.6~\mum feature peaks at a wavelength longwards of 8.62~\mum. Meanwhile, the 6.2~\mum feature peaks between 6.24 and 6.28~\mum, and the 11.2~\mum is relatively broad with a peak at $\sim$11.25~\mum.  In class \classC spectra, the 6.2~\mum feature peaks at 6.29~\mum and the 7-9~\mum complex has a very broad band peaking around 8.2~\mum \citep{peeters2017}. Finally, class \classD has been assigned to PAH spectra with a broad feature peaking at 6.24~\mum, a broad feature ranging from 7-9~\mum peaking at 7.7~\mum, and a peak position at $\sim$11.4~\mum of the 11.2~\mum PAH band  \citep{Matsuura2014, Sloan2014}. It is important to note that PAH spectra do not simply fall into a discrete class, but rather, seem to form a continuous distribution ranging from Class \classA to \classD. 

Observations demonstrate that each class has a connection to the astronomical object in which they are observed. Class \classA band profiles are attributed to interstellar material, typically being observed in reflection nebulae, the ISM, and \HII regions. Class \classB and \classC are associated with circumstellar material, being most commonly detected in post-AGB stars, planetary nebulae, and isolated Herbig AeBe stars (e.g., \citealt{Cohen1989a, Bregman1989, Peeters2002, Tielens2008}).

\section{NGC~7027}
\label{sec:7027}

NGC~7027 is a relatively nearby ($\sim$900 pc) carbon-rich planetary nebula \citep[PN;][]{Masson1989, Ziljstra2008}. At a young dynamical age of roughly 600 years, it is still at a relatively early stage of the PN phase of stellar evolution \citep{Masson1989}. NGC~7027 is very bright; at its centre lies a very hot white dwarf (WD) with an effective temperature of about \mbox{200 000 K} and a luminosity of roughly 8000~$L_{\odot}$ \citep{Latter2000, Zhang2005}. The current core mass is estimated to be $\sim$0.65~$M_{\odot}$ \citep{Ziljstra2008}, while the progenitor mass is estimated to have been 3-4~$M_{\odot}$ \citep{BernardSalas2001}.

\begin{figure}[htbp]
    \centering
    \includegraphics[width=1.\linewidth, clip, trim=3cm 11cm 3cm 15cm]{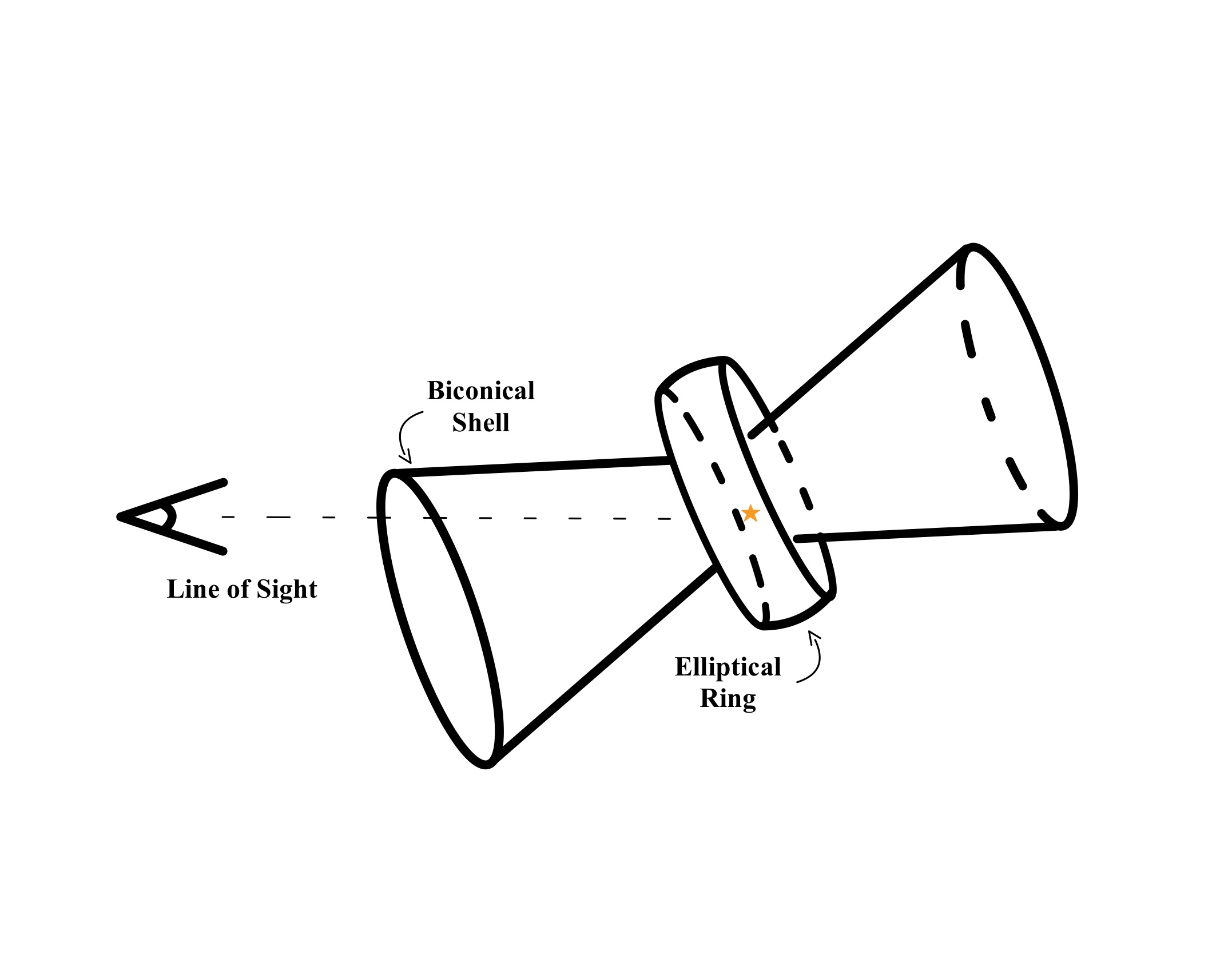}
    \includegraphics[width=.8\linewidth, clip, trim=10cm 34cm 5cm 8cm]{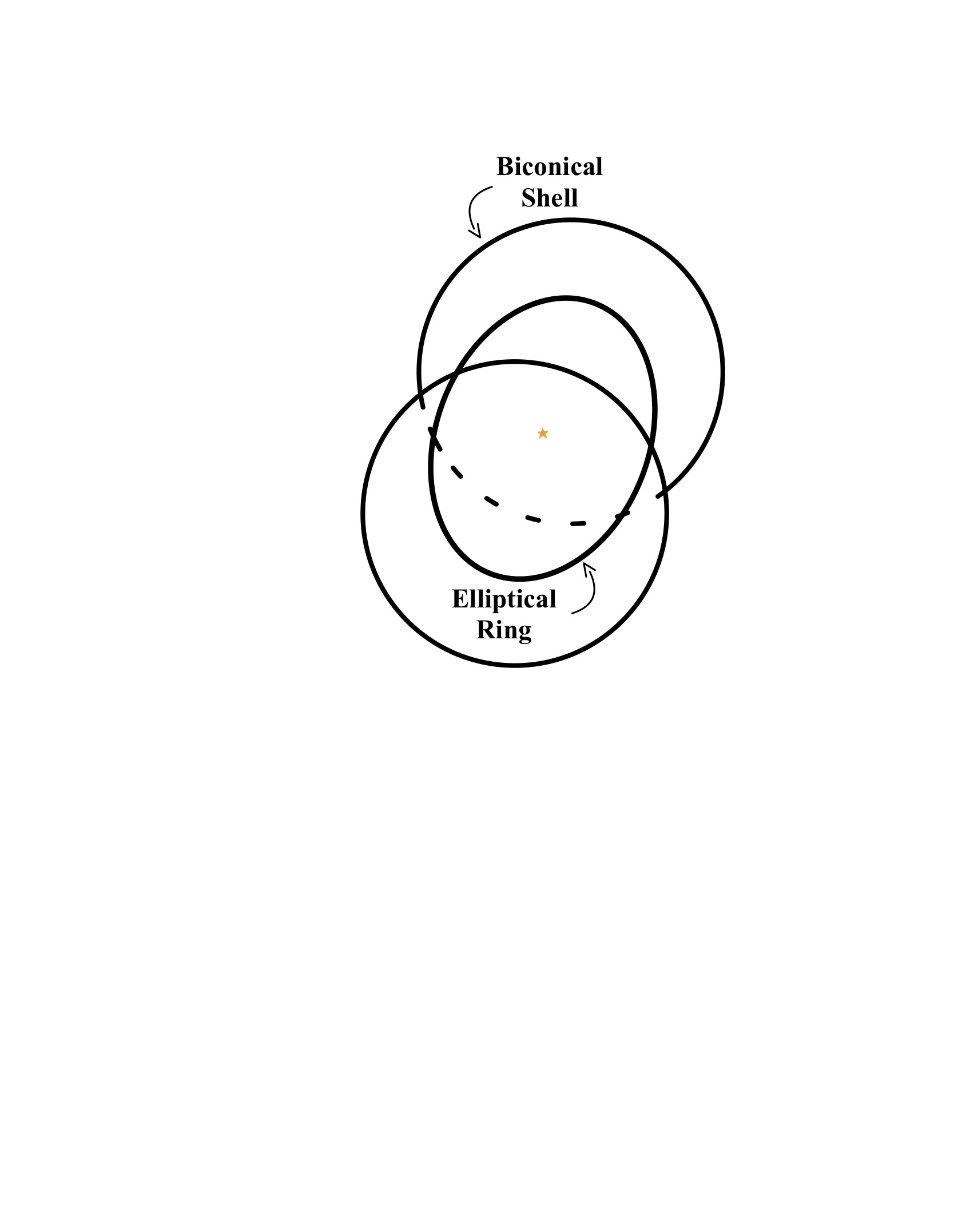}
    \caption{\small Simplified schematic drawings of NGC~7027 as viewed from the side (top) and the front (bottom). The central star in shown in yellow. The \molh emission lies primarily on the surface of the biconical shell, and is particularly bright along the rims, while the ionized gas and PAHs reside within the elliptical ring.}
    \label{fig:schematic}
\end{figure}

Numerous infrared and sub-mm observations have revealed that the material ejected during the AGB phase has formed a flattened ellipsoidal distribution \citep[Fig.~\ref{fig:fov};][]{Cox2002}. This non-spherical distribution is likely due to the presence of a binary system at the centre of the nebula, as many studies of PNe suggest \citep[e.g.,][]{Balick2002, DeMarco2009, MoragaBaez2023}. Strong, precessing jets have carved out bipolar cones that are being illuminated by the EUV and FUV photons from the hot central white dwarf \citep{Latter2000, Cox2002}. These conditions create an ionized nebula, which forms a dense, clumpy, and limb-brightened elliptical ring \citep{Cox2002, Tielens2005}, an adjacent photodissociation region (PDR), and a surrounding molecular cloud, all expanding with a typical velocity of $\sim$20 km s$^{-1}$ \citep{Cox1997, Latter2000, Bublitz2023}. A simplified schematic of NGC~7027 as viewed from the side is shown in Fig. \ref{fig:schematic}.

The \molh morphology is remarkably different from the morphology of the ionized gas. The \molh emission shows a four-lobed structure formed by a limb-brightened biconical shell \citep[Fig.~\ref{fig:fov};][]{Graham1993, Kastner1994}. This hourglass-like shell is thought to be capped at both ends, producing the limb brightening observed in the outer rims of the bicone \citep{Latter2000}. The kinematics of the \molh distribution reveal that the lower cone faces towards the line of sight and has an inclination angle of $i \sim$  60$^\circ$ and the upper cone faces away from the observer \citep{Cox2002}. There is enhanced \molh emission in the equatorial waist due to limb brightening of the inner rims of the biconical shell \citep{Latter2000, Cox2002}. The asymmetry observed in the lobes is due to the presence of three bipolar outflows that, over time, have carved out holes in the \molh distribution, the most pronounced being `Outflow 1' at a position angle of -53$^\circ$ \citep{Cox2002, Nakashima2010}. Appendix~\ref{app:ngc7027_morph} details the \molh, ionized gas, and 6.2~\mum PAH emission as observed by JWST MIRI-MRS within the channel 1 field-of-view (CH1 FOV; Fig.~\ref{fig:3color_images}). The edges of the elliptical ring are pictured by the ionized gas and the 6.2~\mum PAH emission (which sits just outside the \HII region in the PDR), while the equatorial waist is delineated by the \molh emission. The nebula seems to exhibit the typical layered PDR structure. 

The CO emission, which traces the cold, dense material surrounding the PDR, sits further outside the \molh emission and displays a roughly elliptical morphology with particularly bright edges in the NE and SW \citep{Bublitz2023}.

Narrow dust lanes are observed along the equatorial belt, and non-uniform dust filaments are observed surrounding the lobed and elliptical ring structures, most prominently around the northeast lobe \citep{Kastner2020, MoragaBaez2023}. 

Outflow-driven shocks play an important role in the processing of ejected material and are thought to be a mechanism for very small grain and PAH formation in the PN via grain-grain collisions \citep{Lau2016}. 


\section{Observations, data reduction and analysis}
\label{sec:obs_data_reduction}

\begin{figure}[htbp]
    \centering
    \includegraphics[width=.6\linewidth]{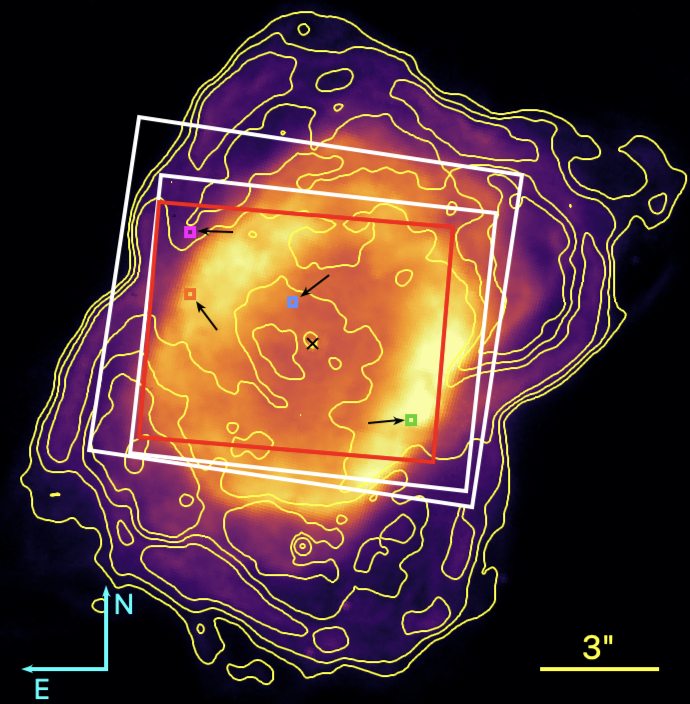}
    \caption{\small HST/WFC3 NICMOS F212N image of NGC~7027 with JWST MIRI-MRS FOVs for channels 1 (red), 2, and 3 (white) overlaid. The black cross indicates the position of the central star. The yellow contours outline the \molh gas morphology in the continuum-subtracted HST/NICMOS images from \cite{Latter2000}. The small orange, pink, blue, and green boxes are the spaxels used to illustrate the AIB variability referred to hereinafter as the E ring, outer NE corner, inner region, and the SW ring, respectively. Credit: NASA/ESA, \cite{Latter2000}.}
    \label{fig:fov}
\end{figure}

We present JWST MIRI Medium Resolution Spectroscopy (MRS) Integral Field Unit \citep[IFU, ][]{MIRI, MRS, MIRIperformance, MRSperformance} observations of NGC~7027 (program ID 01523, observation 1). A 3 $\times$ 3 mosaic was taken centred on the central star. The resulting spectral cubes cover a wavelength range of 4.9--27.9~\mum at a spectral resolution of R$\sim$1500--3500 over a FOV of up to 9.8" $\times$ 8.5" depending on the channel (Fig.~\ref{fig:fov}). 

We reduce the data using the JWST science calibration pipeline version 1.15.1 and the CRDS reference file context 1293. We used the default pipeline settings with the exception of performing a residual fringe correction in stage 2 of the pipeline and setting the output cube types to subchannels. 

We use data from channels 1 to 3 (4.9--18~\mum). For comparative analysis, we use the pixel size of channel 3 (0.2" $\times$ 0.2") and the FOV of the channel 1 mosaic (7.5" $\times$ 6"). We performed multiplicative stitching between subchannels to obtain the final spectra. We analyze individual spaxels and extract a spectrum across the channel 1 FOV, further referred to as the MIRI-MRS integrated spectrum (see Table \ref{tab:aperture_info} for details).

We determine the dust continuum to extract the profiles of the 6.2, 7.7-8.6, and 11.2~\mum PAHs (see Appendix~\ref{app:sec:cont} for details) and classify their profiles based on the classification of \citet{Peeters:prof6:02} and \citet{vanDiedenhoven:chvscc:04}. To facilitate the discussion, we employ four spectra extracted in the apertures shown in Fig.~\ref{fig:fov} (see Table~\ref{tab:aperture_info}), probing the NE corner - outside the elliptical ring (pink), the E side of the elliptical ring (orange), the central region (blue), and the SW portion of the elliptical ring (green). 


\section{Results}
\label{sec:results}

Overall, the PAH emission in the MIRI-MRS integrated spectrum (see Fig.~\ref{fig:miri_integrated_sp_classes}) of NGC~7027 resembles those reported in the literature \citep[e.g., ISO-SWS;][]{Peeters2002, vanDiedenhoven:chvscc:04}. 
However, due to the distinct extraction apertures (7.5"$\times$6" for MIRI-MRS versus 14"$\times$20" for ISO-SWS), differences emerge upon close inspection of the spectral profiles (Fig.~\ref{fig:classes}, Table~\ref{tab:prof_summary}). The MIRI data reveals that the PAH emission varies systematically across the nebula.

\subsection{Spatial Variation of the PAH Spectral Profiles}

\begin{figure*}[htbp]
    \centering
    \resizebox{\hsize}{!}{
    \includegraphics[height=0.3\linewidth, clip, trim=.5cm .9cm .7cm .7cm]{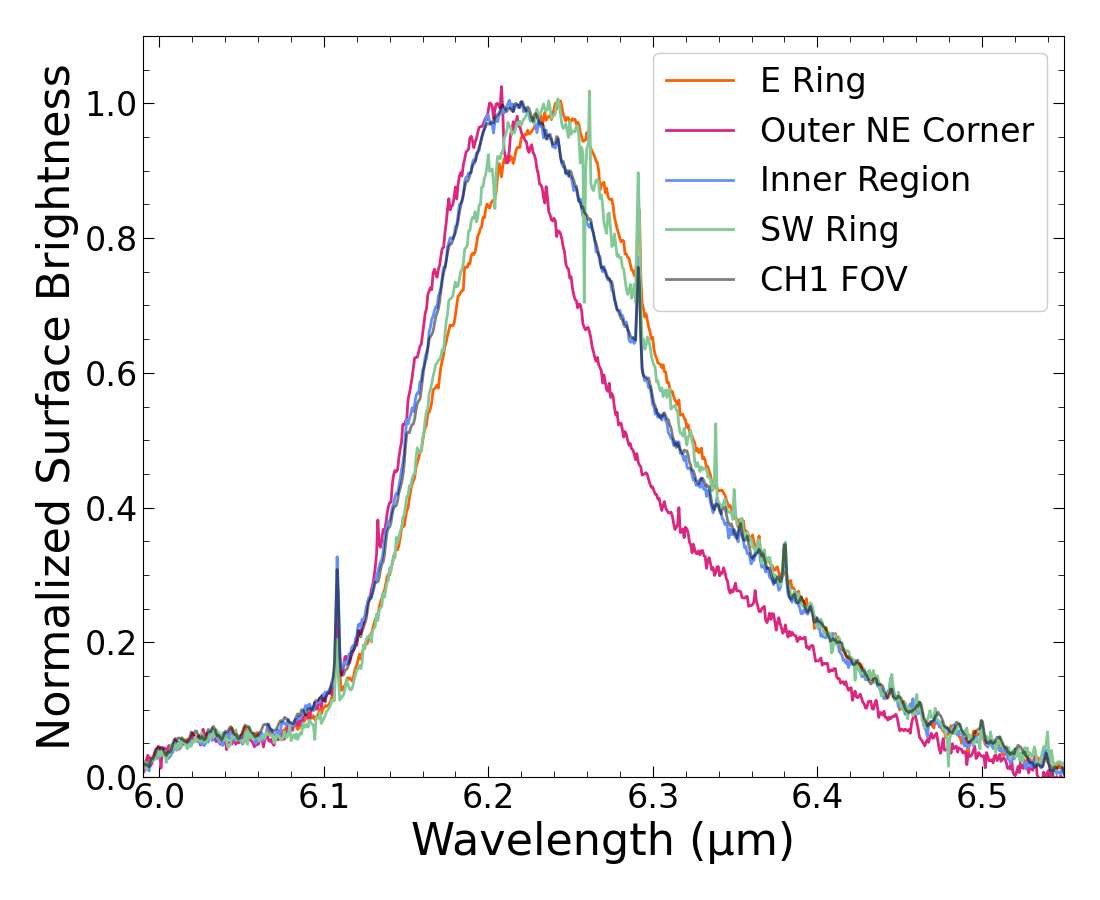}
    \includegraphics[height=0.3\linewidth, clip, trim=.5cm .9cm .7cm .7cm]{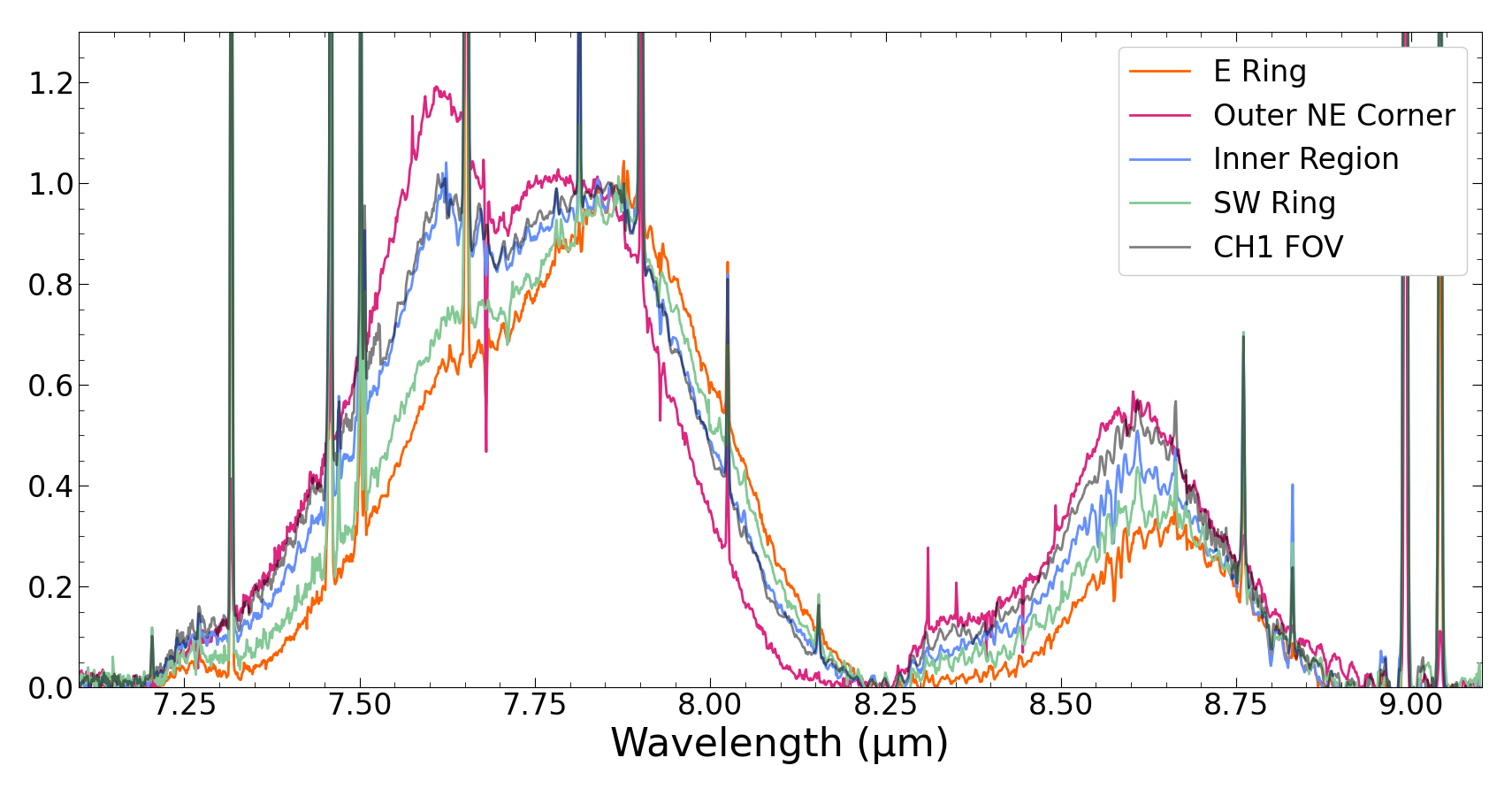}
     \includegraphics[height=0.3\linewidth, clip, trim=.5cm .9cm .7cm .7cm]{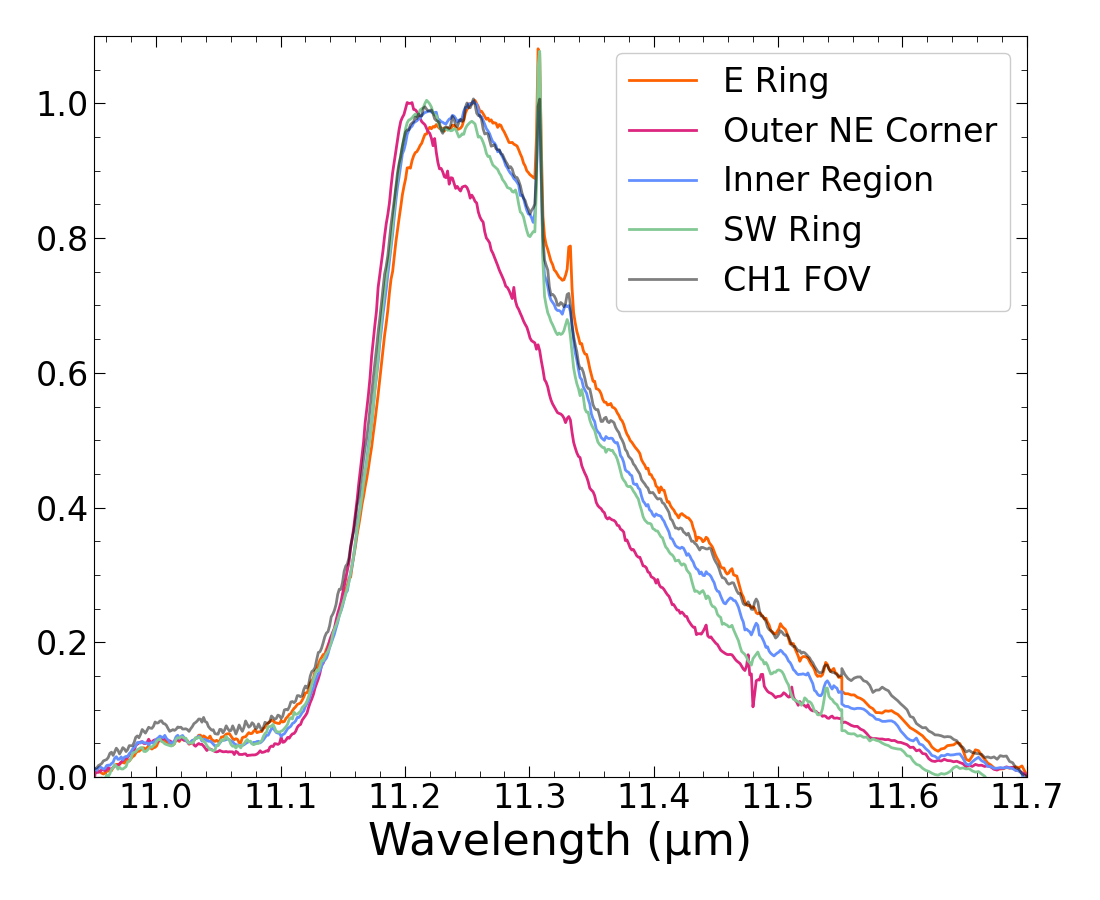}}
   \caption{\small Illustration of the 6.2, 7-9, and 11.2~\mum profile variation across the nebula. The single spaxel spectra from Fig.~\ref{fig:fov} detailed in Table~\ref{tab:aperture_info} are utilized here.}
    \label{fig:classes}
\end{figure*}

The spaxel spectra demonstrate wide PAH profile variation across NGC~7027 in each of the major features observed (Fig.~\ref{fig:classes}). We investigate the profile class of the 6.2, 7.7, 8.6, and 11.2~\mum PAH bands.

The spatial variations in PAH profile classes reveal a striking morphological pattern across NGC~7027. In the 6.2~\mum feature, the outer NE corner of the CH1 FOV (beyond the elliptical ring) is dominated by a strong class~\classA profile, with a narrow profile peaking at 6.205~\mum. Within the ring, however, this shifts to a clear class~\classB profile, broader in shape and peaking at 6.24~\mum. The central region presents an intermediate case: while the peak remains consistent with class~\classA, the band is noticeably broader than in the NE corner.

\begin{figure}[htbp]
    \begin{center}
    \includegraphics[width=.9\linewidth, clip, trim=.8cm 4cm .2cm 0cm]{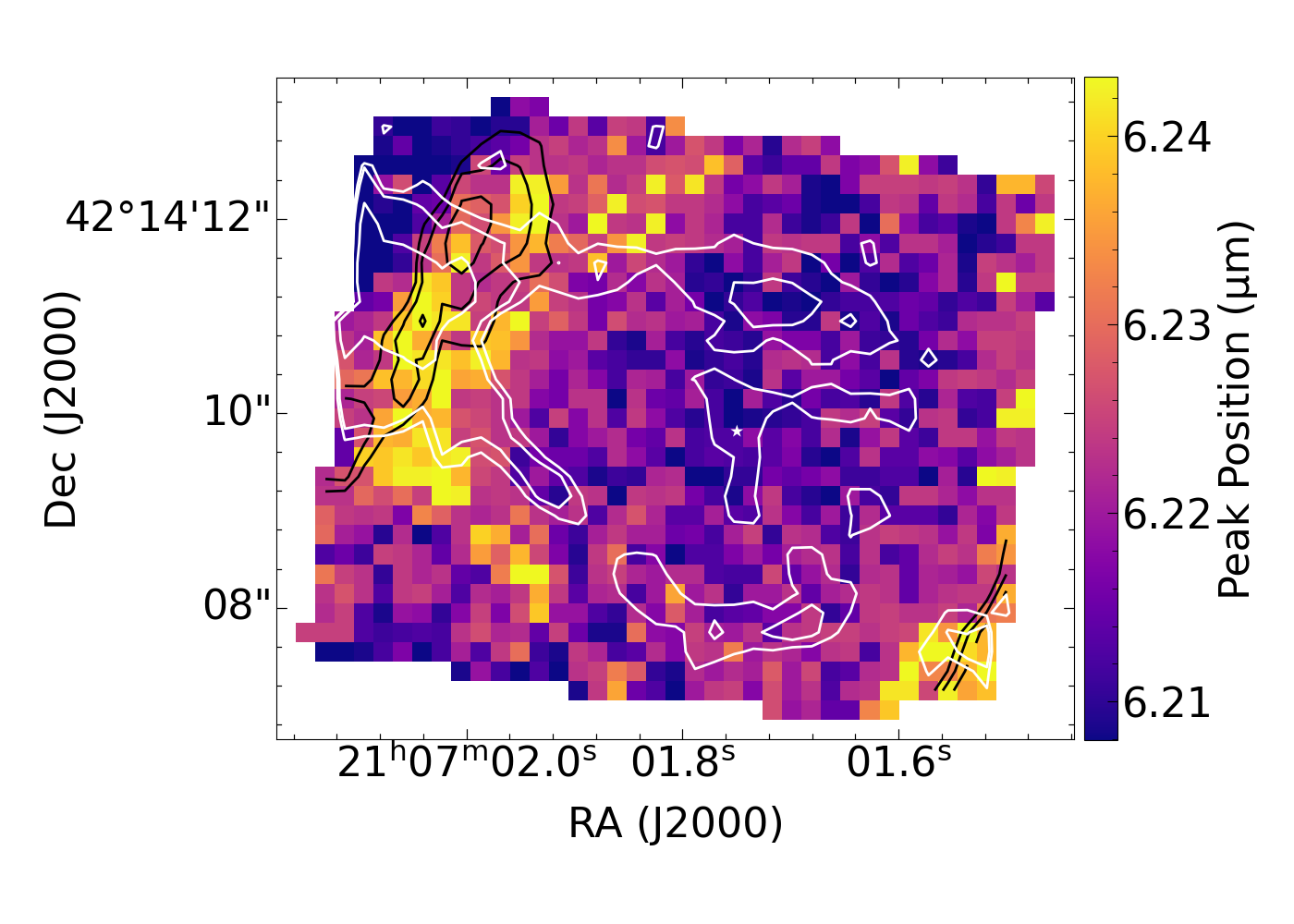}
    \includegraphics[width=.9\linewidth, clip, trim=.8cm 2cm .2cm 1cm]{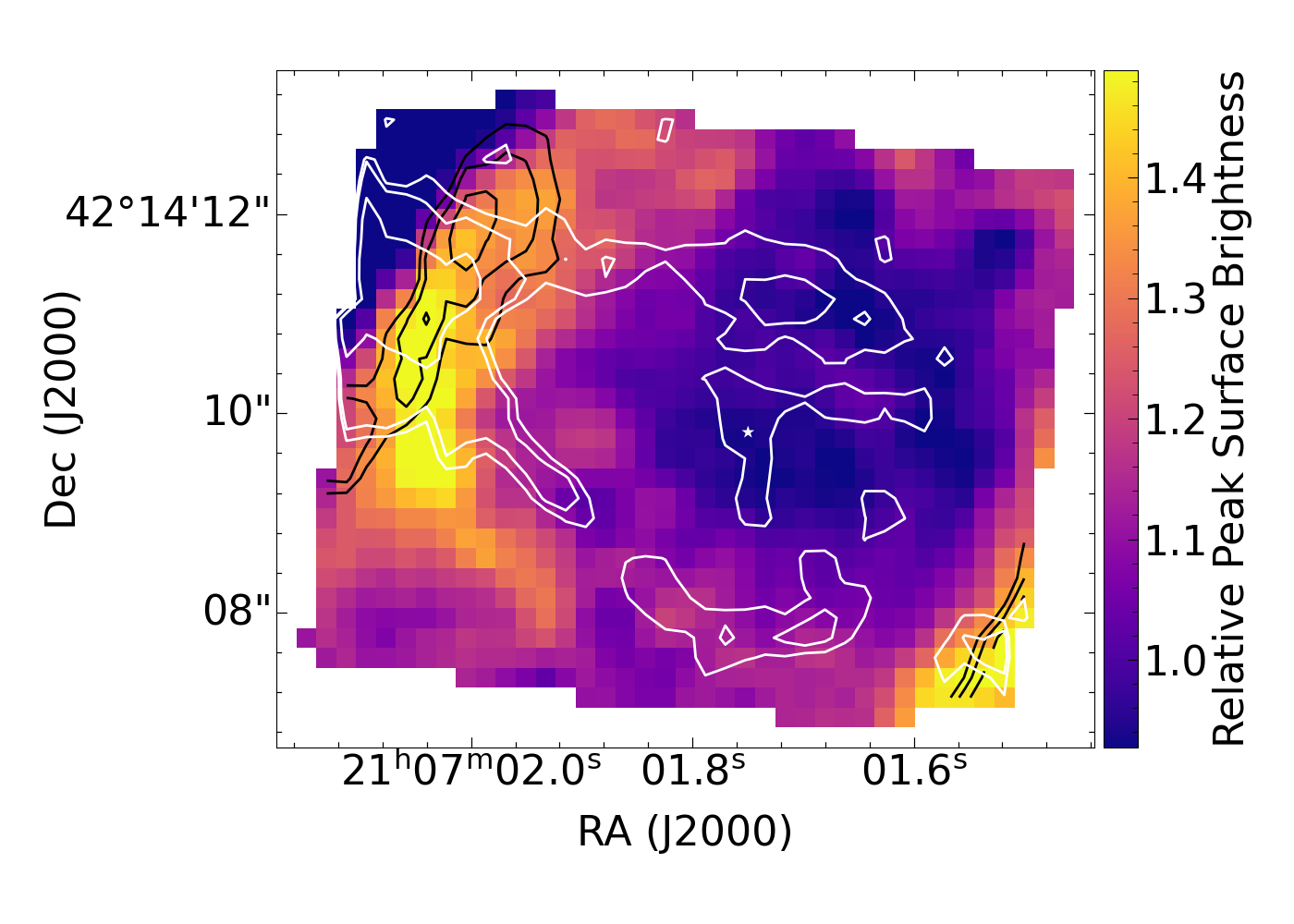}
    \caption{\small Illustration of the profile variability with the peak position of the 6.2~\mum PAH band (top) and the relative peak surface brightness of the 7.8~\mum component with respect to the 7.6~\mum component (bottom). Black and white contours represent the integrated 6.2~\mum PAH surface brightness and the \molh 0-0 S(7) integrated surface brightness, respectively (Fig.~\ref{fig:3color_images}). }
    \label{fig:profile-variability_maps}
    \end{center}
\end{figure}

A similar progression is observed in the 7.7~\mum feature. In the NE corner, a class~\classA profile is evident, where the 7.6~\mum component dominates over the 7.8~\mum band. In contrast, the ring exhibits a pure class~\classB profile with a strong 7.8~\mum peak, while the inner region contains a mixture of intermediate class~\classAB and class~\classB profiles.

The 8.6~\mum feature follows the same spatial trend. A class~\classA profile with a peak at $\sim$8.6~\mum is found in the NE corner, transitioning to a class~\classB profile at $\sim$8.65~\mum within the ring, and reverting to class~\classA profiles in the inner region.

Finally, the 11.2~\mum feature echoes this pattern. The NE corner again shows a class~\classA profile at 11.2~\mum, while the ring hosts class~\classB profiles. The inner region, as with the 7.7~\mum band, displays a mixture of class~\classAB and \classB profiles. Interestingly, scattered class~\classA profiles are detected surrounding the H$_{2}$ emission (Fig.~\ref{fig:fwhm_maps}). We explore the 11.2~\mum band in more detail and its relationship with the \molh emission in the following section.

Summarizing, the spectral profiles of these four PAH bands (6.2, 7.7, 8.6, and 11.2~\mum) display class \classA, \classAB, and \classB profiles which are linked to morphological structures within NGC~7027. 

\subsection{Disentangling the 11.2~\mum Feature}

The 11.2~\mum feature is known to have at least two components, namely the 11.207~\mum and 11.25~\mum \citep{OBpahs}. Originally detected in the Orion Bar, the 11.207~\mum is attributed to PAH emission while the broad 11.25~\mum component is thought to be attributed to very small grains (VSGs) or PAH clusters \citep{OBpahs, OBml, Khan2025}. In the Orion Bar, the 11.25~\mum VSG component is strongest in the \molh dissociation front and beyond. Following the method established by \cite{Khan2025}, we isolated two components at 11.204 and 11.29~\mum, referred to as the 11.207 and 11.25~\mum components, respectively (see Appendix~\ref{app:sec:decompositions} for details).

\begin{figure}[htbp]
    \centering    
    \includegraphics[width=0.8\linewidth, clip, trim=.5cm .7cm .7cm .7cm]{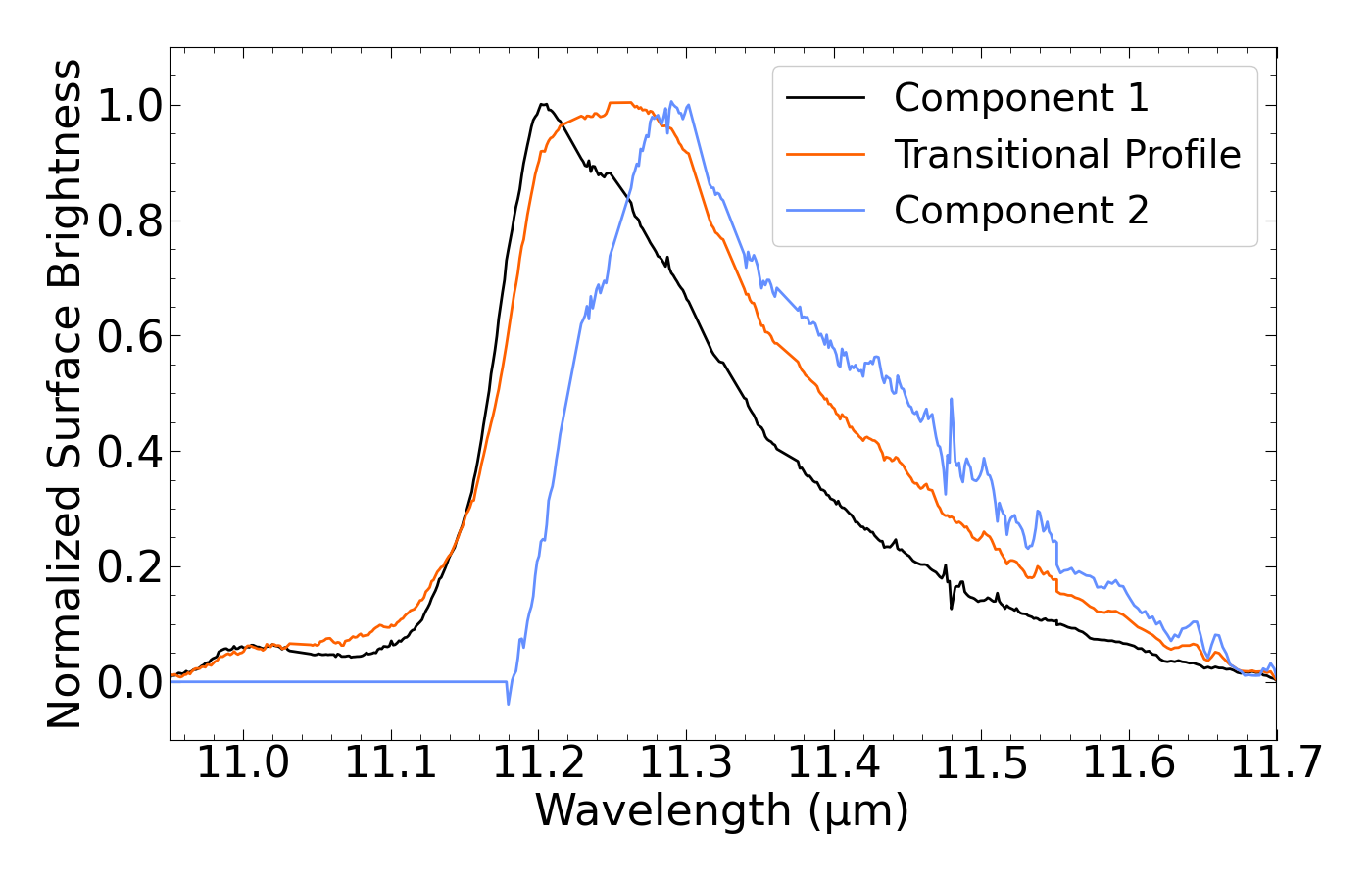}
  \includegraphics[width=0.8\linewidth, clip, trim=.5cm .7cm .7cm .7cm]{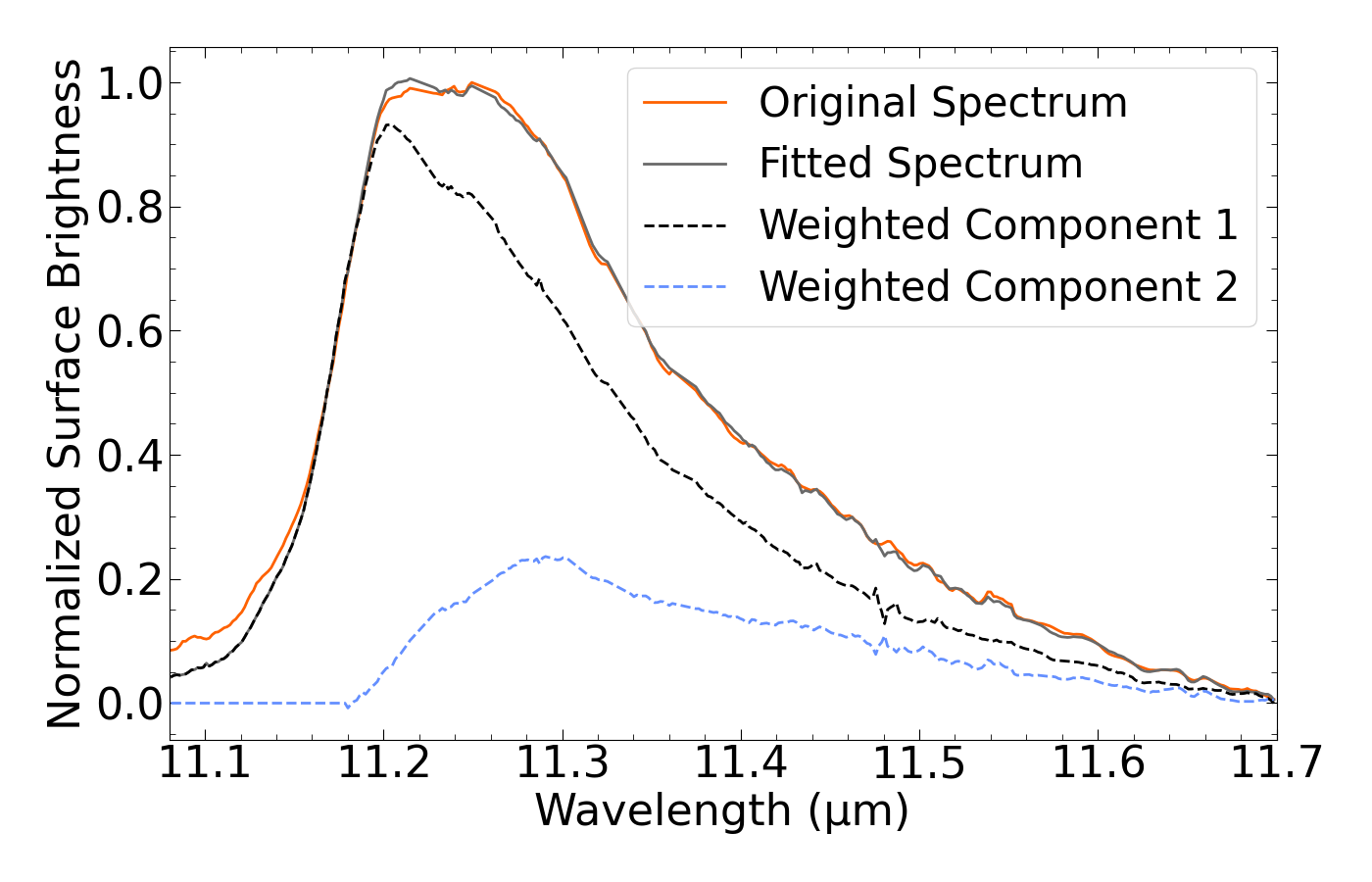}
    \caption{\small Illustration of the 11.2~\mum PAH band decomposition. Top: Normalized profiles of the first component, the narrowest profile, representing class \classA, the broadest (transitional) profile representing class \classB, and the second component used in the decomposition of the 11.2~\mum feature. Bottom: Typical decomposition of the 11.2~\mum band by components 1 and 2.}
    \label{fig:112comps}
\end{figure}

\begin{figure}[htbp]
    \centering    
       \resizebox{.7\hsize}{!}{\includegraphics[width=0.4\linewidth, clip, trim=2cm 2cm 1cm .7cm]{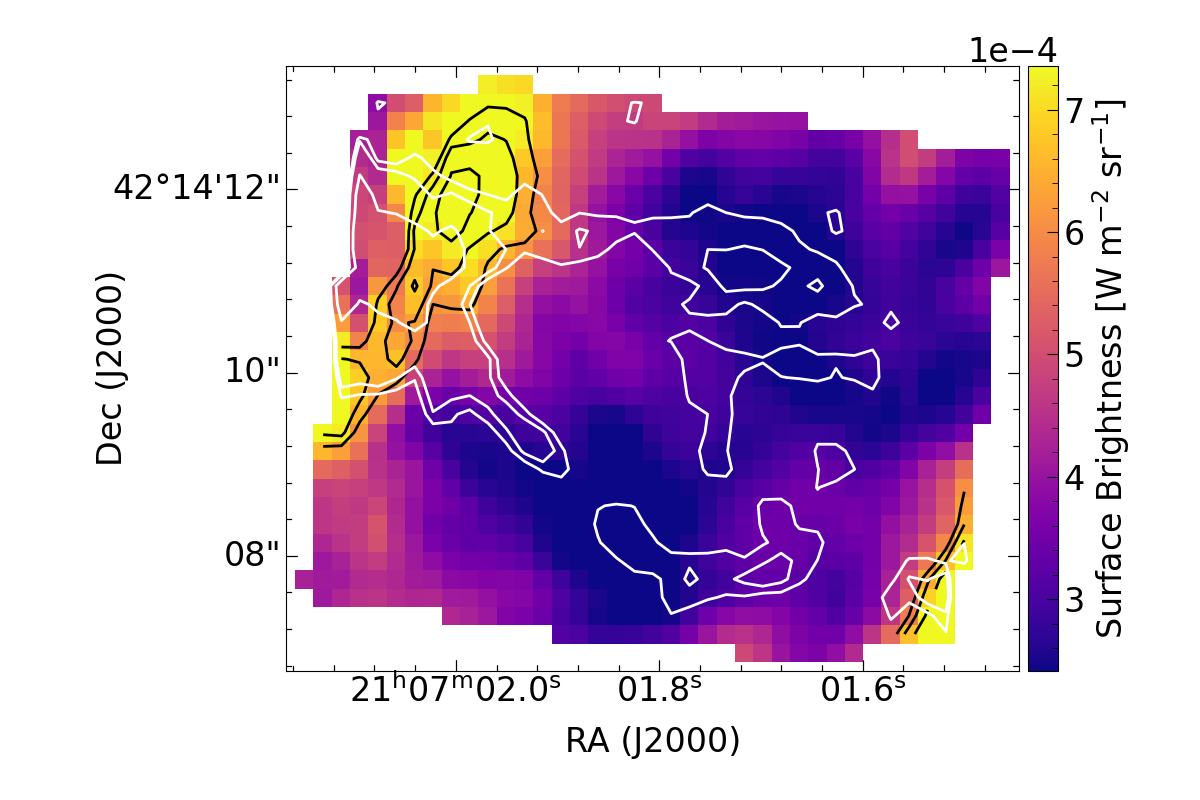}}
          \resizebox{.7\hsize}{!}{\includegraphics[width=0.5\linewidth, clip, trim=2cm 1cm 1cm .7cm]{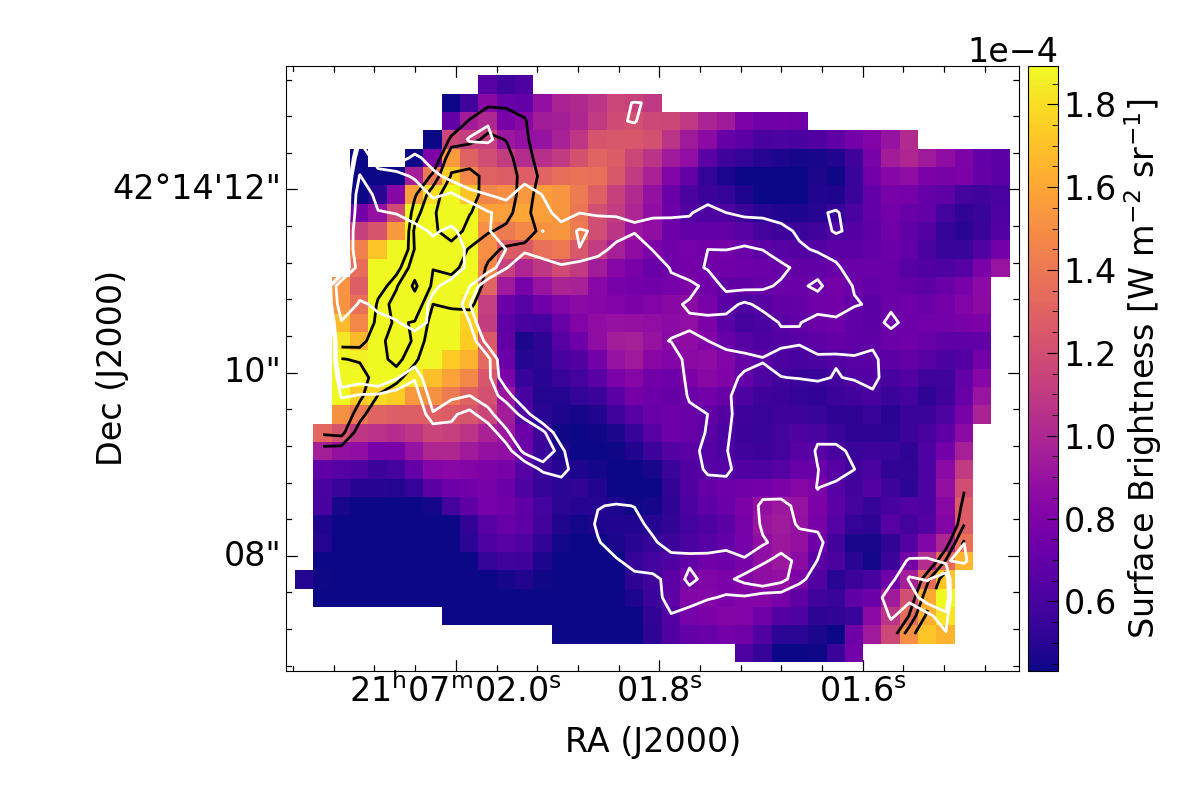}}
    \caption{\small Spatial maps of the integrated surface brightness of the 11.207~\mum component (top) and of the 11.25~\mum component (bottom). Contours are as in Fig.~\ref{fig:profile-variability_maps}.
    }
    \label{fig:112_comp_map}
\end{figure}

Spatial mapping (Fig.~\ref{fig:112_comp_map}) reveals that the 11.207~\mum component has a morphology akin to that of the 6.2~\mum PAH feature, while the 11.25~\mum component has a spatial distribution roughly resembling the \molh morphology, similar to the Orion Bar observations, reinforcing that the 11.25~\mum component is indeed distinct from the PAH emission. 

\subsection{Correlation Studies of the 6-9~\mum Features}

\begin{figure}[htbp]
    \centering   
    \includegraphics[width=0.8\linewidth, clip, trim=.5cm .7cm .7cm .7cm]{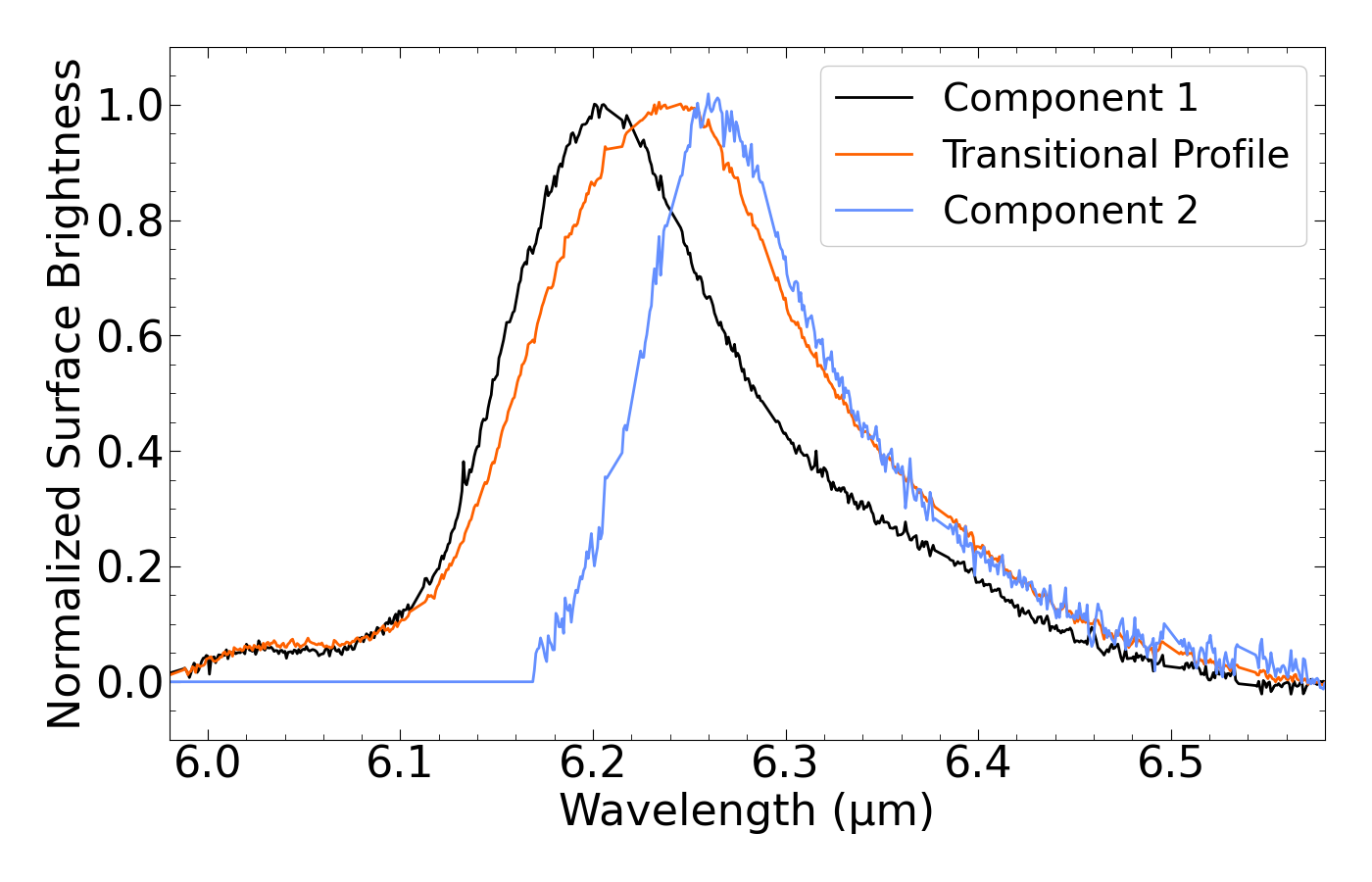}
 \includegraphics[width=0.8\linewidth, clip, trim=.5cm .7cm .7cm .7cm]{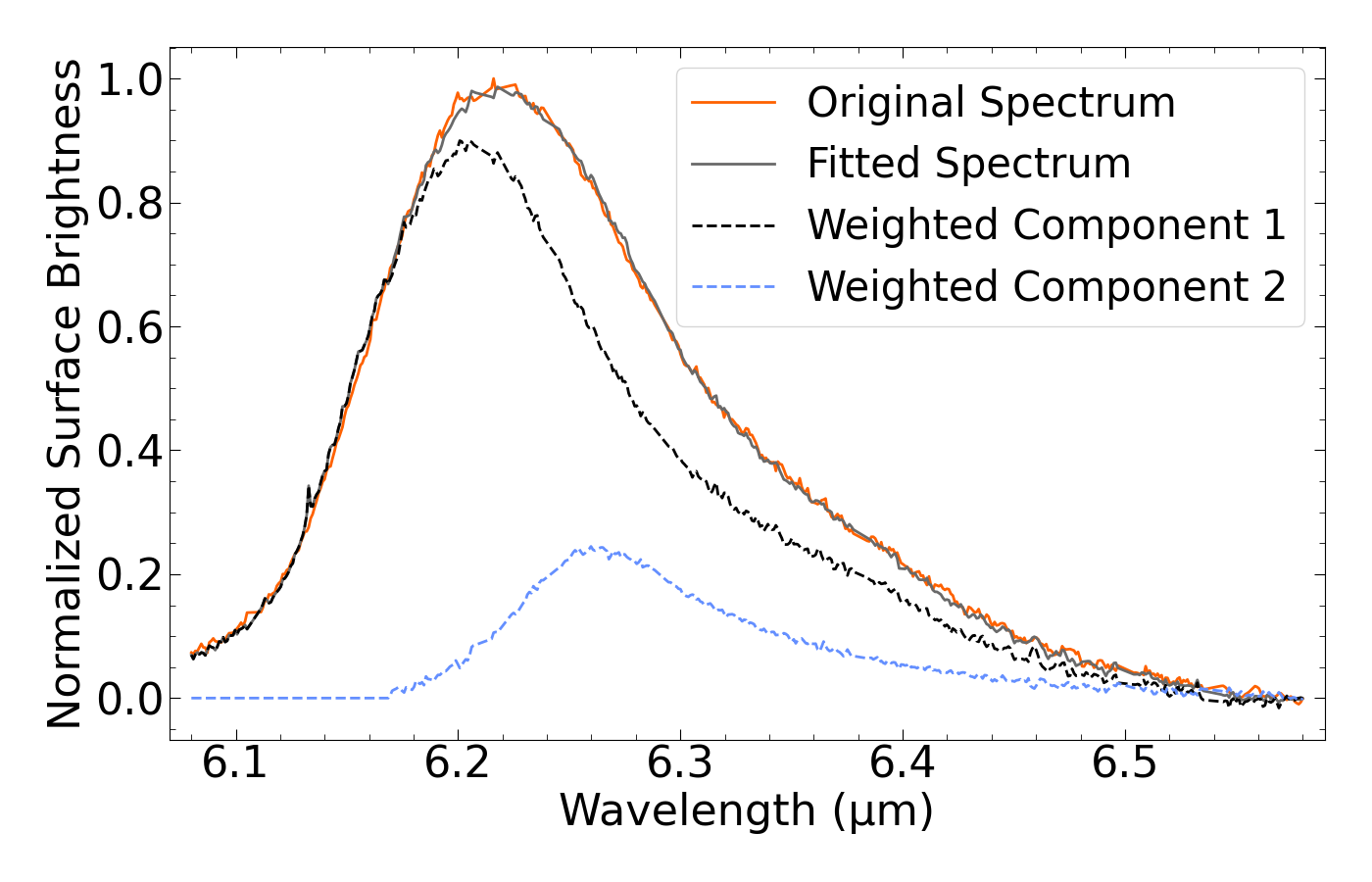}
    \caption{\small Illustration of the decomposition of the 6.2~\mum PAH band. Top: Normalized profiles of the first component, the narrowest profile representing class \classA, the broadest (transitional) profile representing class \classB, and the derived second component used in the decomposition. Bottom: Typical decomposition of the 6.2~\mum PAH band with components 1 and 2.}
    \label{fig:62comps}
\end{figure}

\begin{figure}[htbp]
    \centering   
 \includegraphics[width=0.8\linewidth, clip, trim=.5cm .7cm .7cm .7cm]{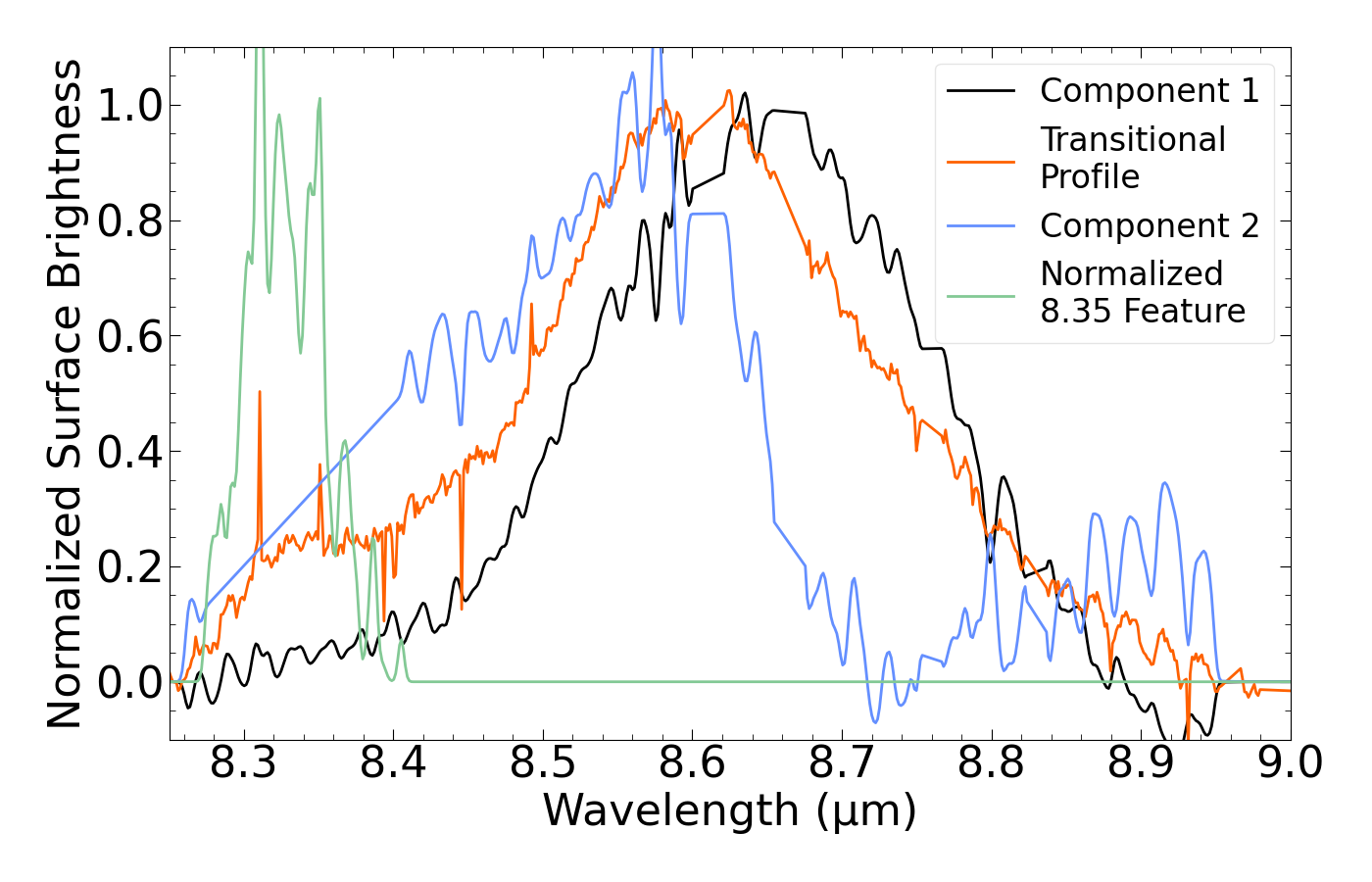}
 \includegraphics[width=0.8\linewidth, clip, trim=.5cm .7cm .7cm .7cm]{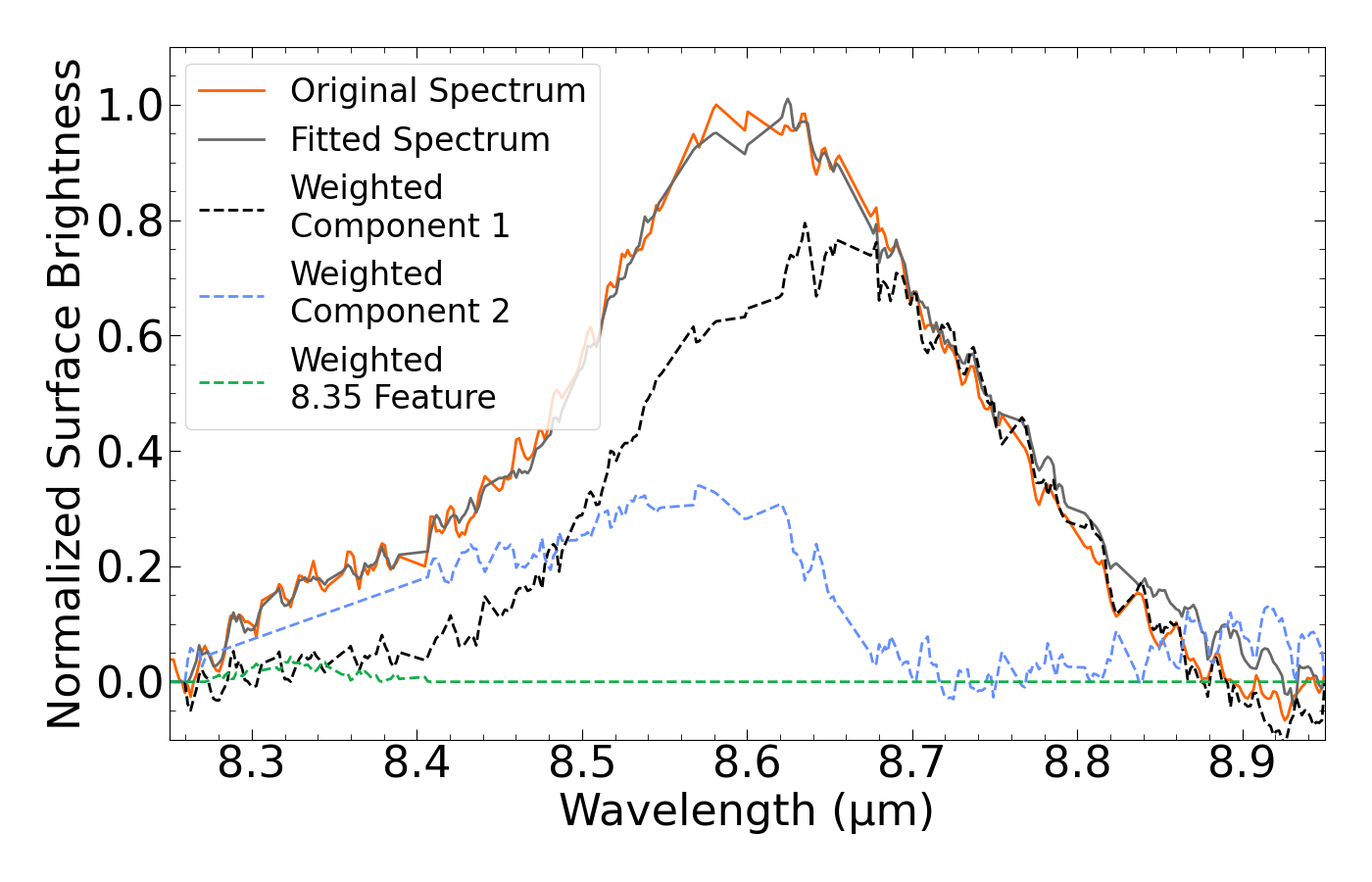}
    \caption{\small Illustration of the decomposition of the 8.6~\mum PAH band. Top: Normalized profiles of the red-shifted profile representing class \classB (the first component), the blue-shifted profile representing class \classA (the transitional profile), the derived second component used in the decomposition, and the extracted 8.35~\mum feature. Bottom: Typical decomposition of the 8.6~\mum PAH band with components 1 and 2, and the 8.35~\mum feature.}
    \label{fig:86comps}
\end{figure}

Motivated by the significant profile variation in the 6.2 and 8.6~\mum features, we perform a similar decomposition as was done for the 11.2~\mum band to investigate whether the variation arises from blending of two components. For the 6.2~\mum feature, two components at 6.205~\mum and 6.26~\mum are successfully extracted and fit to reproduce all spectra (Fig.~\ref{fig:62comps}). The 6.205~\mum component is the narrowest profile observed whilst the 6.26~\mum component is extracted via subtraction of the 6.205~\mum component from broadest profile observed. Fig.~\ref{fig:86comps} shows two components successfully isolated from the 8.6~\mum feature, at 8.56 and 8.65~\mum. Here, the most red-shifted 8.6~\mum profile is assigned as the first component and subtracted from the 8.6~\mum profile exhibiting the most blue-shift to extract the second component. The weak 8.35~\mum feature is extracted from the second component to isolate its emission. Details of the decompositions are found in Appendix~\ref{app:sec:decompositions}.

As the 7.6 and 7.8~\mum components in the 7.7~\mum feature trace class \classA versus class \classB PAH populations, a decomposition was performed to examine the spatial distribution of each component independently. To do so, we fit the 7-9~\mum complex using four Gaussian components representing the 7.6, 7.8, 8.2, and 8.6~\mum features, consistent with the approach of \cite{peeters2017} and \cite{Stock2017}, but adopting a broader 8.2~\mum component. The data are well-modelled with these Gaussian components and the integrated surface brightnesses of each component are measured (see Appendix~\ref{app:sec:decompositions} for details).

\begin{figure*}[htbp]
    \centering
    \includegraphics[width=\linewidth, clip, trim=0cm .5cm 0cm .5cm]{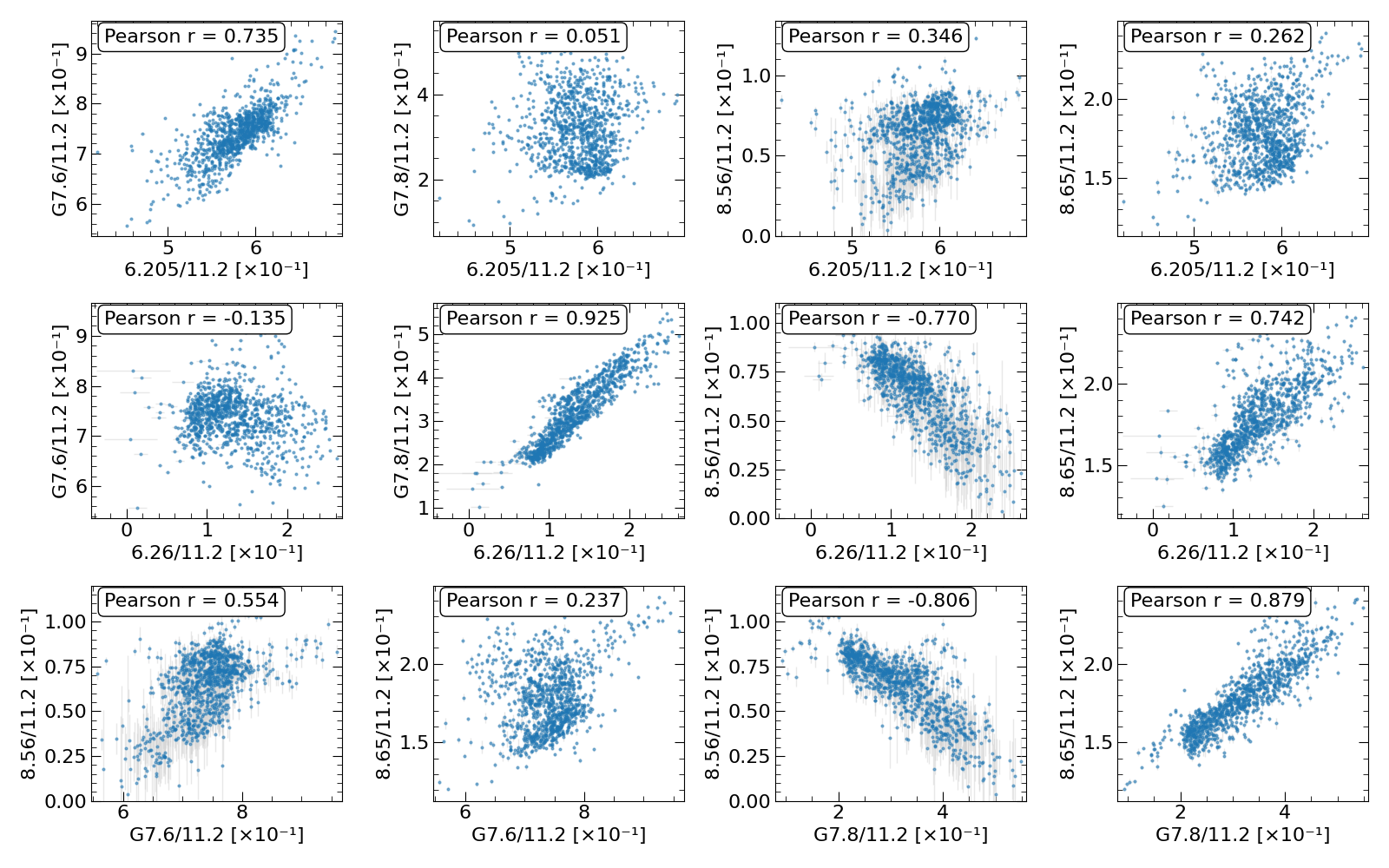}
    \caption{\small Correlations between the 6.205, 6.26, G7.6, G7.8, 8.56, and 8.65~\mum integrated component surface brightnesses normalized over the total 11.2~\mum integrated surface brightness. Correlation coefficients are shown at the top and error bars are shown in light grey.}
    \label{fig:62_77_86_ComponentCorrelations}
\end{figure*}

Analysis of component correlations provides further insight into the PAH interrelationships. Correlation plots between the 6.2, 7.7, and 8.6~\mum components reveal that the blue components of the 6.2 and 7.7~\mum bands (6.205 and G7.6~\mum) are strongly correlated but neither are closely correlated with the blue or red component of the 8.6~\mum band (Fig.~\ref{fig:62_77_86_ComponentCorrelations}).  The three red components of the 6.2, 7.7, and 8.6~\mum features (6.26, G7.8, and 8.65~\mum) show tight correlations with each other, but none demonstrate correlations with the 6.205 and G7.6~\mum components. The 8.56~\mum component demonstrates an anti-correlation with the red components of the 6.2 and 7.7~\mum components and is, therefore, separate from all other components. We refer the reader to Fig.~\ref{fig:heatmap} for the correlation coefficients between all 6-9~\mum components.

Examining component ratios (Fig.~\ref{fig:comp_correlations_norm_over_total}) further highlights the uniqueness of the 11.25~\mum component. The fractional contribution of the red component to the 6.2 (6.26/6.2), 7.7 (G7.8/7.7), 8.6 (8.65/8.6) ratios correlate strongly with each other, yet none correlate with the fractional contribution of the VSG component to the 11.2 (11.25/11.2), reinforcing the conclusion that the 11.25~\mum emission arises from a carrier distinct from the carriers of the blue or red components of the 6.2, 7.7, and 8.6~\mum PAH bands. 

We thus find four distinct PAH subpopulations: the red 6-9~\mum components (6.26, 7.8, 8.65~\mum); the blue 6.205 and 7.6~\mum components; the blue 8.56~\mum component; and the 11.207~\mum neutral PAH component. Additionally, the VSG/PAH cluster population is represented by the 11.25~\mum component.

\section{Discussion}
\label{sec:discussion}

\subsection{Spectral friends}
\label{disc:spectral_friends}

The correlation observed between the 6.2 and 7.7~\mum blue components (6.205~\mum and 7.6~\mum) and the red components (6.26~\mum, 7.8~\mum, and 8.65~\mum) of the 6.2, 7.7, and 8.6~\mum PAH bands (Sec.~\ref{sec:results}) suggests that these components originate from two distinct PAH subpopulations, with a separate third subpopulation represented by the blue 8.56~\mum component of the 8.6~\mum feature. Understanding the underlying origin of these spectral relationships is crucial.

\begin{figure}[htbp]
    \centering
    \includegraphics[width=\linewidth]{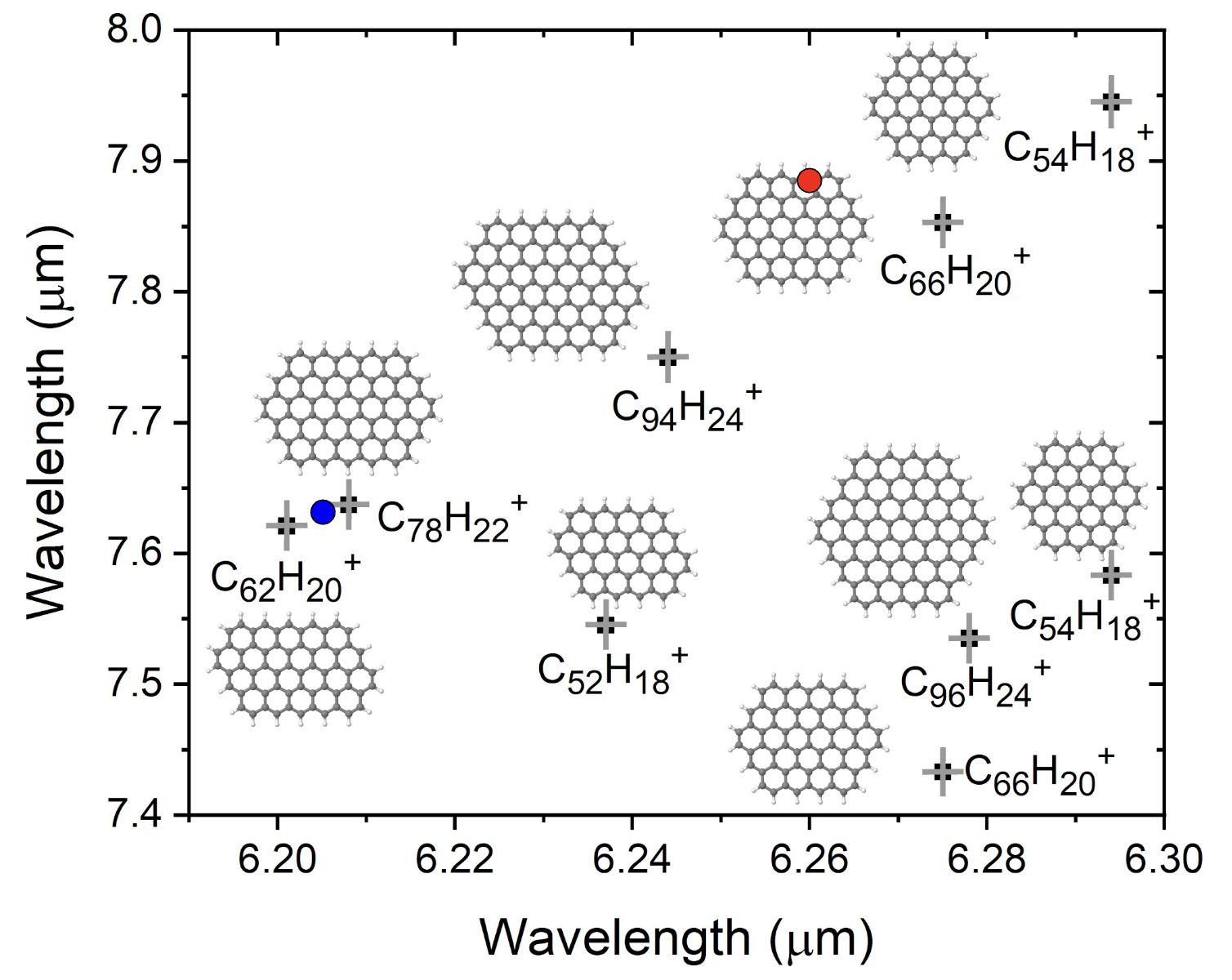}
    \caption{\small The 6.2 and 7.7~\mum peak positions of 7 cationic PAH species \cite[data taken from][]{Ricca2021}. Uncertainties for the theoretical band positions, calculated in \cite{Mackie2016, Mackie2018} and \cite{Lemmens2019}, are shown by the grey error bars (for details see Appendix~\ref{app:sec:band_positions}). The blue and red circles represent the observed peak positions of the blue and red components of the 6.2 and 7.7~\mum bands in NGC~7027, respectively.}
    \label{fig:PAHspecies_62_77_PeakPos}
\end{figure}

\citet{Ricca2021} reported that the 6.2~\mum feature peaks at bluer wavelengths in elongated PAHs and at redder wavelengths in circular PAHs. Expanding on this result, Fig.~\ref{fig:PAHspecies_62_77_PeakPos} illustrates the 6.2 and 7.7~\mum peak positions for a set of PAH molecules, offering insight into potential carriers of these emission components. We use the NASA Ames PAH Infrared Spectroscopic Database\footnote{\url{https://www.astrochemistry.org/pahdb}} version 4.0 \cite[PAHdb;][Ricca et
al. under review]{Bauschlicher:10, Boersma2014, Bauschlicher2018, Mattioda2020} to relate the peak positions of the 6.2 and 7.7~\mum bands to their corresponding molecular structures for a collection of seven PAH species. Only cationic PAHs were selected because the 6.2 and 7.7~\mum bands are dominated by PAH cations \citep{Hudgins1999a}. We further constrained the subset of species by limiting ourselves to PAHs with long straight edges containing adjacent C-H solo bonds (also denoted as zigzag edges) because the 11.2~\mum feature, attributed to the solo C-H out-of-plane bending mode, dominates the 10-15~\mum C-H out-of-plane bending modes features, indicating that compact species dominate the PAH family \citep{hony2001, Ricca2018, Ricca2021}. 

We find that the PAHdb calculations of the 6.2 and 7.7~\mum peak positions of these species span the full wavelength range covered by the 6.2 and 7.7~\mum components in NGC~7027 and that the precise positions of these features are unique for specific PAH molecules. If these findings are confirmed by quantum chemistry studies on a larger set of astrophysically relevant species, the peak positions of the components contributing to the 6.2 and 7.7~\mum features could prove to be a useful diagnostic of PAH molecules in space. Based on this small sample, species with a similar structure to C$_{62}$H$_{20}$$^{+}$ and C$_{78}$H$_{22}$$^{+}$ appear to be good candidates for the 6.205 and 7.6~\mum components and molecules analogous to C$_{66}$H$_{20}$$^{+}$ are potential carriers of the 6.26 and 7.8~\mum components. We do not exclude, however, the possibility that each subpopulation originates not from a single PAH species but is instead made up of a group of PAH species sharing similar spectral characteristics whose modes blend together into a single component. We do realize that the photochemical behaviour of these putative species will have to be very similar as well to preserve the good correlations observed while the PAH population is being processed by the strong radiation field.

Previous studies link these PAH components to structural characteristics. For instance, \cite{peeters2017} attributed the G7.6~\mum component to compact, cationic, small PAHs and the G7.8~\mum component to neutral, very large PAHs or PAH clusters. However, our results here are inconsistent with these findings. We instead see the G7.8/G7.6~\mum ratio map (Fig.~\ref{fig:78over76_ratio_map}) closely resembles the 6.2/11.207 ratio map (Fig.~\ref{fig:62over11207_ratio_map}), a known tracer of the ionic to neutral PAH ratio. This suggests instead that the conditions that favor the G7.8~\mum component also favor charged PAHs, while the conditions that favor neutral PAHs also favor the G7.6~\mum component. However, the C-C stretching modes are very weak in neutral PAHs \citep{Hudgins1999a} and such species are not expected to contribute major features to the 6-9~\mum range \citep{Oomens2003, Bauschlicher:2008}. It is likely that these variations have no connection with PAH charge but instead with both PAH size and structure
\citep{Bauschlicher:2008, Bauschlicher2009}. It has been suggested that a red peak position of the 7.7~\mum band is indicative of PAHs with $\sim$96-130~C \citep{Langhoff:96, Bauschlicher:2008, Joblin2008, Bauschlicher2009}, albeit a few smaller sized PAHs also show a red 7.7~\mum band (see e.g., Fig~\ref{fig:PAHspecies_62_77_PeakPos}). We do not see such a trend in the small subset of PAHs included in Fig.~\ref{fig:PAHspecies_62_77_PeakPos} due to its limited sample and size range. 

Density Functional Theory (DFT) studies imply that the intrinsic strength of the 8.6~\mum in-plane C-H bending mode is a strong function of the PAH size and molecular structure \citep{Bauschlicher:2008, Bauschlicher2009, ricca2012}. Comparison of these quantum chemistry studies with observations imply that the 8.6~\mum band is carried by a subset of compact PAHs with sizes in the range $\sim$96-150 C atoms. Irregular or smaller compact PAHs do not show a very distinct 8.6~\mum band while very large compact PAHs ($>$150 C atoms) show an 8.6~\mum band that is much too strong compared to the C-C modes. 

Given the 6.205 and 7.6~\mum components have no clear 8.6~\mum counterpart, this serves as an indication that these blue components originate from medium-sized, asymmetric PAHs, consistent with \citet{Ricca2021}. However, medium-sized asymmetric PAHs may be carriers of the weak 8.35~\mum feature since these PAHs have intrinsically weak C-H in-plane bending modes. Unfortunately, as the 8.35~\mum feature is a weak band perched on the side of the much stronger 8.6~\mum feature, it is difficult to ascertain a strong link between the 8.35~\mum feature and the 6.205 and 7.6~\mum components. The subpopulation traced by the red components (6.26, 7.8, and 8.65~\mum) is likely composed of compact PAHs within the $\sim$96-150 C-atom size range, which is set by the 8.65~\mum band. We note that the PAHs shown in Fig.~\ref{fig:PAHspecies_62_77_PeakPos} exhibit C-H in-plane bending emission shortwards of 8.5~\mum. The anti-correlation observed between the 8.56~\mum component and the red components of the 6.2 and 7.7~\mum features indicate that the same conditions that favour the 6.205 and 7.6~\mum, hamper the 8.56~\mum. The lack of correlation with any other component suggests that perhaps the 8.56~\mum component stems from PAHs that are very large ($>$150 C atoms). In very large PAHs, the positive charge per carbon is very small and they behave more like neutrals, which have insignificant 6.2 and 7.7~\mum emission \citep{ricca2012}.

Since the 3.3/11.2 PAH band ratio is a known diagnostic of PAH size (e.g., \citealt{Allamandola1989, Pech2002, ricca2012}), future observations of NGC~7027 including the 3.3~\mum feature would prove useful in linking the PAH size across the nebula to the variations found in this analysis of the 6.2, 7.7, and 8.6~\mum features. 

\subsection{PAH clusters \& VSGs - Identity crisis of class \classB$_{11.2}$}

The 6.2, 7.7, and 8.6~\mum PAH bands exhibit class \classA, \classAB, and \classB profiles across the nebula. However, the variation in the 11.2~\mum band is dominated by the contribution from the newly identified 11.25~\mum component, extracted here for the first time in a PN. As was demonstrated in Sec.~\ref{sec:results}, this component behaves distinct from the PAH emission. Our results are consistent with the assignment of the 11.25~\mum emission to PAH clusters and/or very small grains (VSGs) put forward by \citet{OBpahs}, \citet{OBml}, and \citet{Khan2025}.  Consequently, it is difficult to interpret the variation in the 11.2~\mum feature as a shift in PAH spectral class since the 11.25~\mum component overshadows any subtle profile variations towards class \classB. In fact, this component calls into question the existence of the 11.2~\mum class \classB profile defined in \cite{vanDiedenhoven:chvscc:04}, as it is tarnished by the VSG contribution. The unique behaviour of the class \classB$_{11.2}$ was already noted in \cite{vanDiedenhoven:chvscc:04}, where it was observed that class \classB$_{11.2}$ differs from class \classB$_{6-9}$. The recent discovery of the 11.25~\mum component perhaps provides an explanation for the anomalous behaviour of the \classB$_{11.2}$. 

\subsection{PAH evolution}
\label{disc:pah_evolution}

The spatial distribution of PAH spectral classes observed within the nebula (Sec.~\ref{sec:results}) has the potential to provide valuable insights into the evolutionary processing mechanisms acting upon PAHs. Class \classB profiles appear predominantly in the elliptical ring, while class \classA profiles appear in the region external to the ring. We propose that the primary mechanism responsible for processing these PAHs is the star's intense UV radiation \citep[e.g.,][]{Cox2002, MoragaBaez2023}. Consequently, these results suggest that the class \classB PAHs are the material processed by UV radiation, while the class \classA PAHs are unrefined and shielded from such energetic interactions. This is a slightly different scenario than originally suggested by \citet{Peeters2002} in which PAHs are injected by PNe into the ISM as class \classB and the conversion to class \classA occurs in the ISM, driven by the harsh radiation field. We emphasize, though, that the $5-15$ $\mu$m spectra analyzed here are very consistent with predominantly aromatic materials (cf., Fig.~\ref{fig:PAHspecies_62_77_PeakPos}). Near-infrared observations may provide insight into whether the photochemically-driven loss of small PAHs in NGC~7027 is accompanied by a loss of aliphatic material as seen in some post-AGB objects \citet{goto2003,goto2007}.
 
We reconcile our results with earlier studies where class \classA is observed in the ISM and class \classB PAHs in PNe and protoplanetary disks as follows. In PNe, most of a star's initial mass is expelled during the AGB phase. The majority of this ejected material will never be processed by the star's harsh UV radiation or shocks, preserving its original, pristine state as it migrates into the ISM. PAHs in the ISM will, therefore, retain their original class \classA spectral profiles, consistent with PAH observations in the ISM. Conversely, despite the cooler temperatures in protoplanetary disk environments, the class \classB profiles observed could be linked to the strong radiation fields incident on these disks, effectively processing the PAH population. 

Theoretical studies show that shocks can also process the circumstellar and interstellar PAH family \citep{Micelotta2010}. The circumstellar environment of NGC~7027 is heavily processed by the jet(s) emanating from the star(s) \citep{Lau2016, Bublitz2023}. However, the regions most heavily processed fall outside of the FOV of these observations. Future observational studies could be instrumental in testing these models and elucidating the effects of shocks on the PAH family in space.

\section{Conclusions}
\label{sec:conclusions}

We present, for the first time, high-resolution spectral data of the mid-IR PAH emission in NGC~7027 from JWST MIRI-MRS spectral cubes. Our findings are as follows:

\begin{enumerate}
    \item We report the first observation of PAH spectral profile changes over different classes (here \classA, \classAB, and \classB) in each of the major features observed (6.2, 7.7, 8.6, and 11.2~\mum) within a single extended target. 
    \item We decompose the 6.2~\mum feature into two new  components at 6.05 and 6.26~\mum, the 7.7~\mum feature into two Gaussian components: G7.6 and G7.8~\mum, and the 8.6~\mum feature into two components at 8.56 and 8.65~\mum. Correlation plots reveal that the 6.05 and G7.6~\mum components are strongly correlated and the red components of the three features (6.26, G7.8, and 8.65~\mum) exhibit a tight correlation, while no correlation is observed between the red and blue components of the features. The 8.56~\mum component is not observed to have a strong relationship with any of the other components. We, therefore, hypothesize that these components represent three distinct PAH subpopulations, with the blue 6.2 and 7.7~\mum components belonging to 
    medium-sized asymmetric PAHs (like e.g. C$_{62}$H$_{20}^+$), the red 6.2, 7.7, and 8.6~\mum components to large (96-150 C atoms), compact symmetric PAHs and the 8.56~\mum component to very large PAHs ($\ge$150 C atoms).
    \item A successful decomposition is performed of the 11.2~\mum band into its two formerly reported components: 11.207 and 11.25~\mum \citep{OBml, Khan2025}. We confirm that these components behave similarly in NGC~7027 as in previous studies of the Orion Bar, in which the 11.207~\mum component represents the neutral PAH emission and the 11.25~\mum component represents unique emission potentially from VSGs or PAH clusters, thus, casting doubt on the existence of the class \classB 11.2~\mum profile as described by \cite{vanDiedenhoven:chvscc:04}. The 11.2~\mum band shows no clear counterpart to the red/blue components observed in the 6.2, 7.7, and 8.6~\mum bands, nor is any link found between these components and the 11.25~\mum component.
    \item Class \classB profiles are observed in the elliptical ring closer to the harsh UV radiation whereas class \classA profiles are detected further outside the ring. The profile variations across NGC~7027, therefore, indicate that class \classA PAH profiles presumably originate from less refined PAHs and that the class \classB PAH emission arises from more processed PAH species. This implies that the majority of PAHs present in PNe enter the ISM as class \classA, shifting the perception of the evolution of PAH profile classes throughout the PAH lifecycle.
\end{enumerate}

Future spectroscopic work needs to be done to investigate the origin of the different PAH feature components and profile classes. Additional high-resolution spectral data from NGC~7027 and other PNe would help in disentangling the nature of PAHs in the early phases of their evolution.

\newcommand{\dataset}[2][]{#2}

\begin{acknowledgements}
This work is based on observations made with the NASA/ESA/CSA James Webb Space Telescope. The data were obtained from the Mikulski Archive for Space Telescopes at the Space Telescope Science Institute, which is operated by the Association of Universities for Research in Astronomy, Inc., under NASA contract NAS 5-03127 for JWST. These observations are associated with program \#1523 (DOI: 10.17909/5kp7-y840).

Els Peeters acknowledges support from the Natural Sciences and Engineering Research Council of Canada.

Alessandra Ricca acknowledges support from the Internal Scientist Funding Model (ISFM) Laboratory Astrophysics Directed Work Package at NASA Ames (22-A22ISFM-0009).

This article is based upon work from COST Action CA21126 - Carbon molecular nanostructures in space (NanoSpace), supported by COST (European Cooperation in Science and Technology).  
\end{acknowledgements}

\bibliographystyle{aa}
\bibliography{bib}

\begin{appendix}

\section{NGC~7027 Morphology}
\label{app:ngc7027_morph}

\begin{figure}[htbp]
    \centering
   \resizebox{\columnwidth}{!}{
    \includegraphics[width=\linewidth, clip, trim=1.2cm .4cm 4cm 1.8cm]{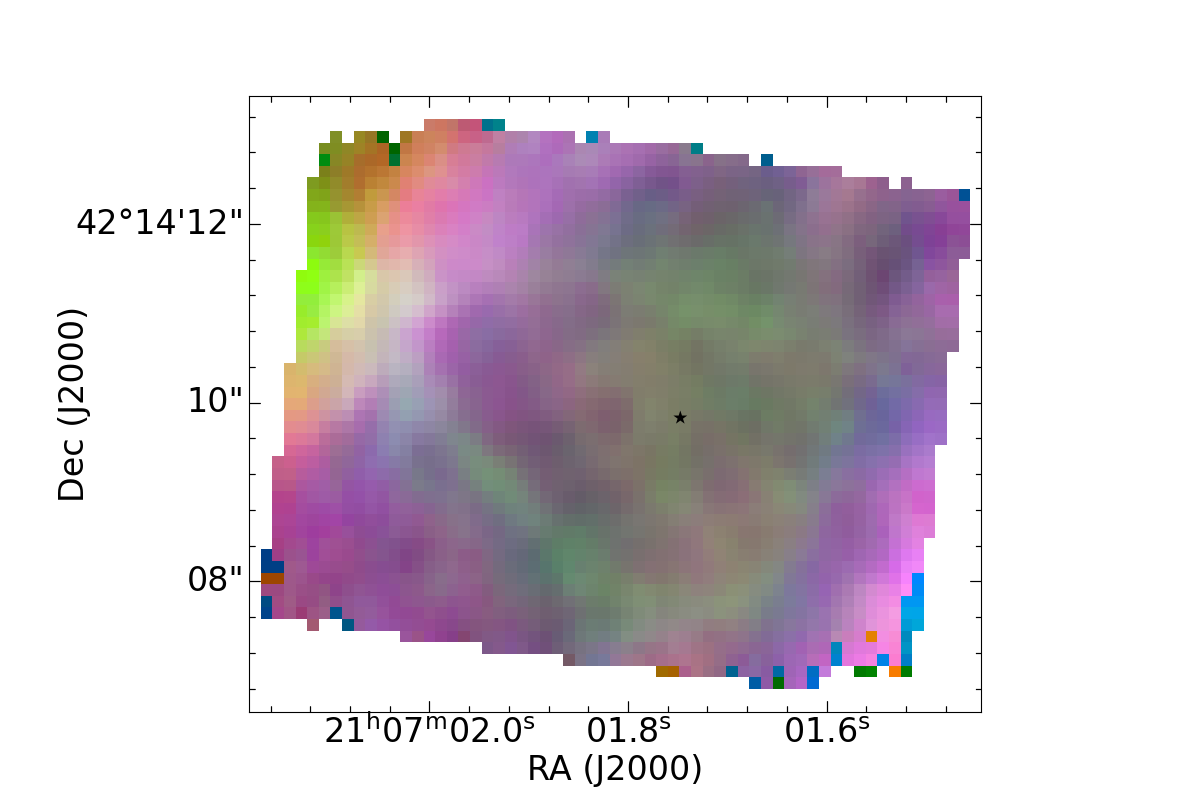}
    }
     \resizebox{\columnwidth}{!}{\includegraphics[width=.0391\linewidth, clip, trim=0.2cm .2cm 27.3cm 1.5cm]{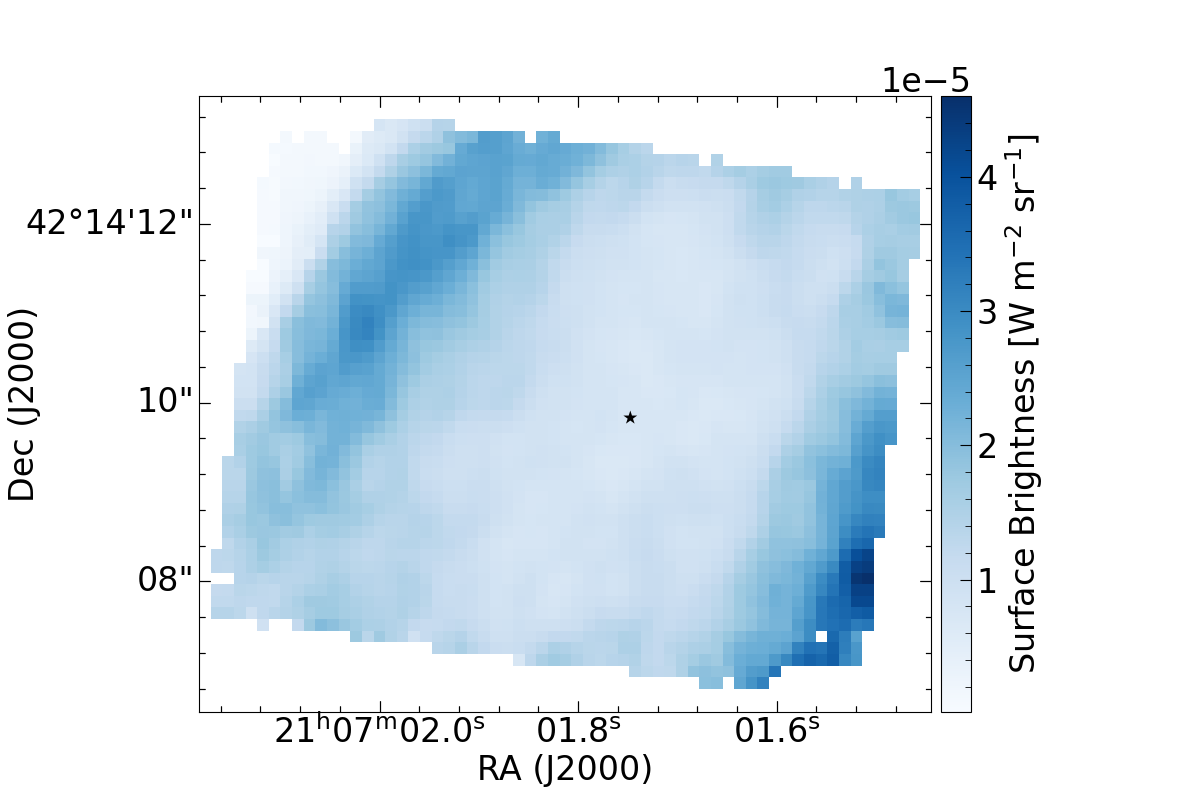}
    \includegraphics[width=.321\linewidth, clip, trim=3.55cm .2cm 2.5cm 1.5cm]{Figures/HI_blue.png}
    \includegraphics[width=.321\linewidth, clip, trim=3.55cm .2cm 2.5cm 1.5cm]{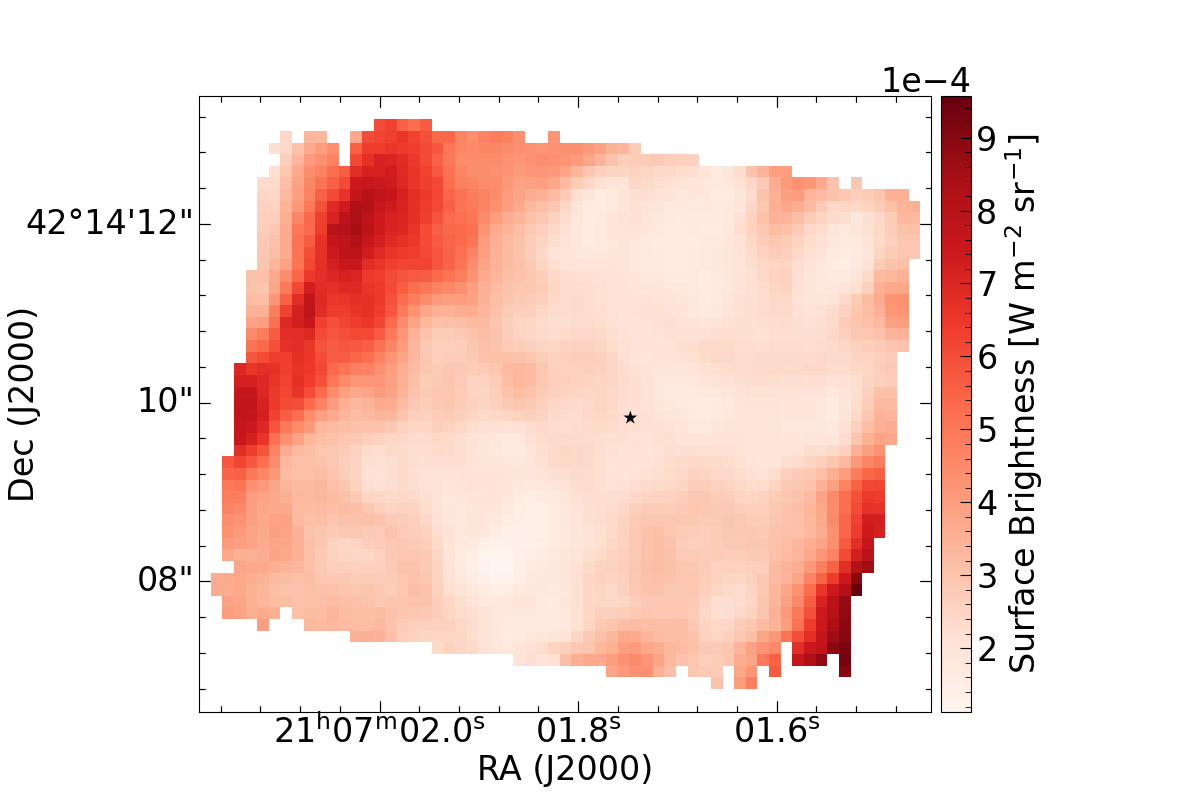}
    \includegraphics[width=.321\linewidth, clip, trim=3.55cm .2cm 2.5cm 1.5cm]{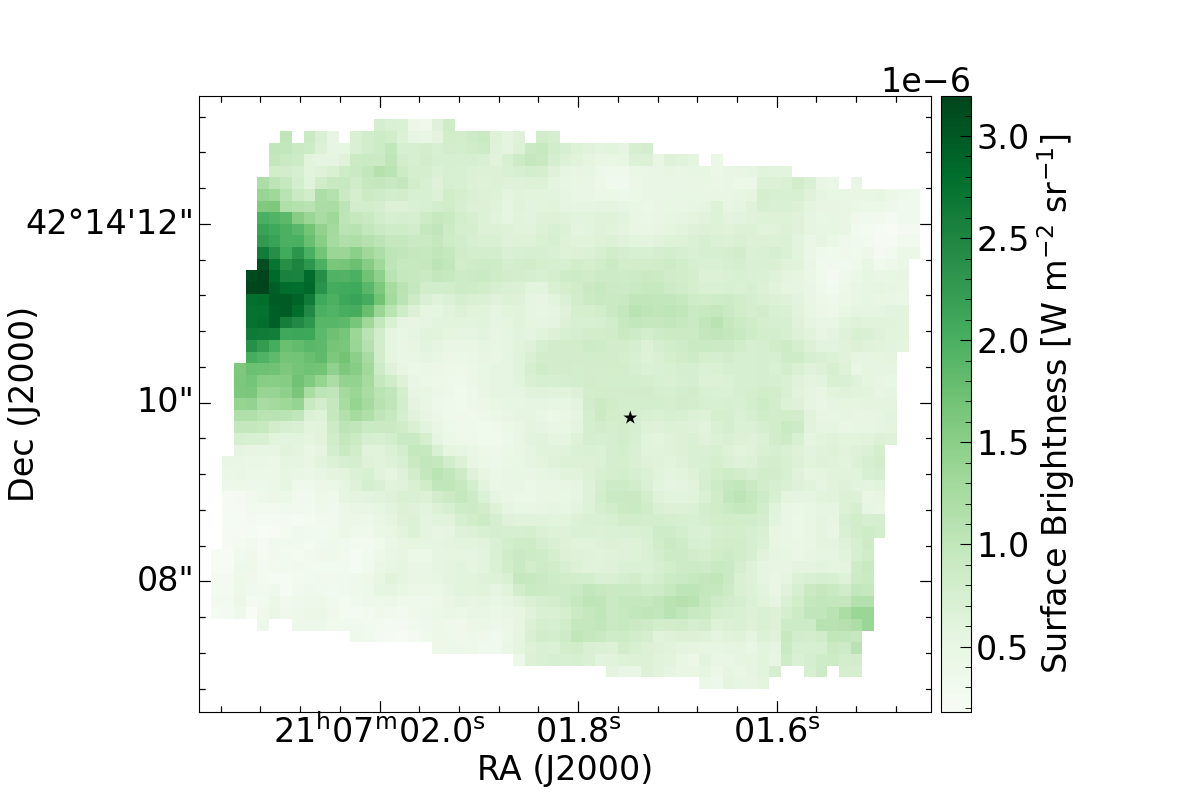}}
    
    \caption{\small Three-colour image of the 6.2~\mum PAH feature (red), \molh 0-0 S(7) emission line (green), and the Pfund $\alpha$ (n=6-5) emission line (blue) integrated surface brightness RGB maps in the NGC~7027 MIRI CH1 FOV (top). The individual integrated surface brightness images are shown in the bottom panels. }
\label{fig:3color_images}
\end{figure}

The morphological structures in the MIRI MRS CH1 FOV of NGC~7027 are demonstrated by the integrated surface brightness maps of the \HI recombination line Pfund $\alpha$ (n=6-5), 6.2~\mum PAH feature, and the \molh 0-0 S(7) emission line, shown in Fig.~\ref{fig:3color_images}. The Pfund $\alpha$ emission traces the \HII region which has the form of an elliptical shell. The 6.2~\mum PAH emission sits just outside of the Pfund $\alpha$ emission and traces the atomic PDR, and morphologically resembles the ionized gas. Finally, the \molh emission outlines the surface of the molecular PDR.

\section{Extraction Apertures}

We utilize five extraction apertures in this study for the MIRI-MRS spectral cube and reference the ISO-SWS data. The information about the position and size each of these is given in Table \ref{tab:aperture_info}. The MIRI-MRS data have a position angle of 74.67$^\circ$.

\begin{threeparttable}[]
    \centering
        \caption{\small Information about the NGC~7027 extraction apertures utilized. Right ascension and declination (J2000) are given in [21:07:ss.ss] and [+42:14:ss.ss] respectively.}
    \label{tab:aperture_info}
    \begin{tabular}{lllc}
    \hline
    \multicolumn{1}{c}{Aperture} & \multicolumn{1}{c}{RA} & \multicolumn{1}{c}{Dec} & \multicolumn{1}{c}{Size} \\
    & (ss.ss) & (ss.ss) & \multicolumn{1}{c}{("$\times$")}\\
\hline\\[-7pt]

    ISO-SWS FOV$^1$ & 01.70 & 09.10 & 14$\times$20 \\[2pt]
    
    MIRI-MRS CH1 FOV & 01.802 & 10.01 & 7.5$\times$6.0 \\[2pt]

    Outer NE Corner Pixel & 02.044 & 12.56 & 0.2$\times$0.2 \\[2pt]
  
    E Ring Pixel & 02.044 & 10.96 & 0.2$\times$0.2 \\[2pt]

    Inner Region Pixel & 01.810 & 10.76 & 0.2$\times$0.2 \\[2pt]

    SW Ring Pixel & 01.540 & 07.76 & 0.2$\times$0.2 \\[2pt]

    \hline \\[-7pt]
 
    \end{tabular}
    \begin{tablenotes}
\item [1] Target Dedicated Time (TDT) number is 55800537.
\end{tablenotes}
\end{threeparttable}

\section{Continuum determination}
\label{app:sec:cont}

\begin{figure}[htbp]
    \centering
    \resizebox{.8\columnwidth}{!}{\includegraphics[width=\linewidth, clip, trim=1.5cm 0cm 2.cm 1.5cm]{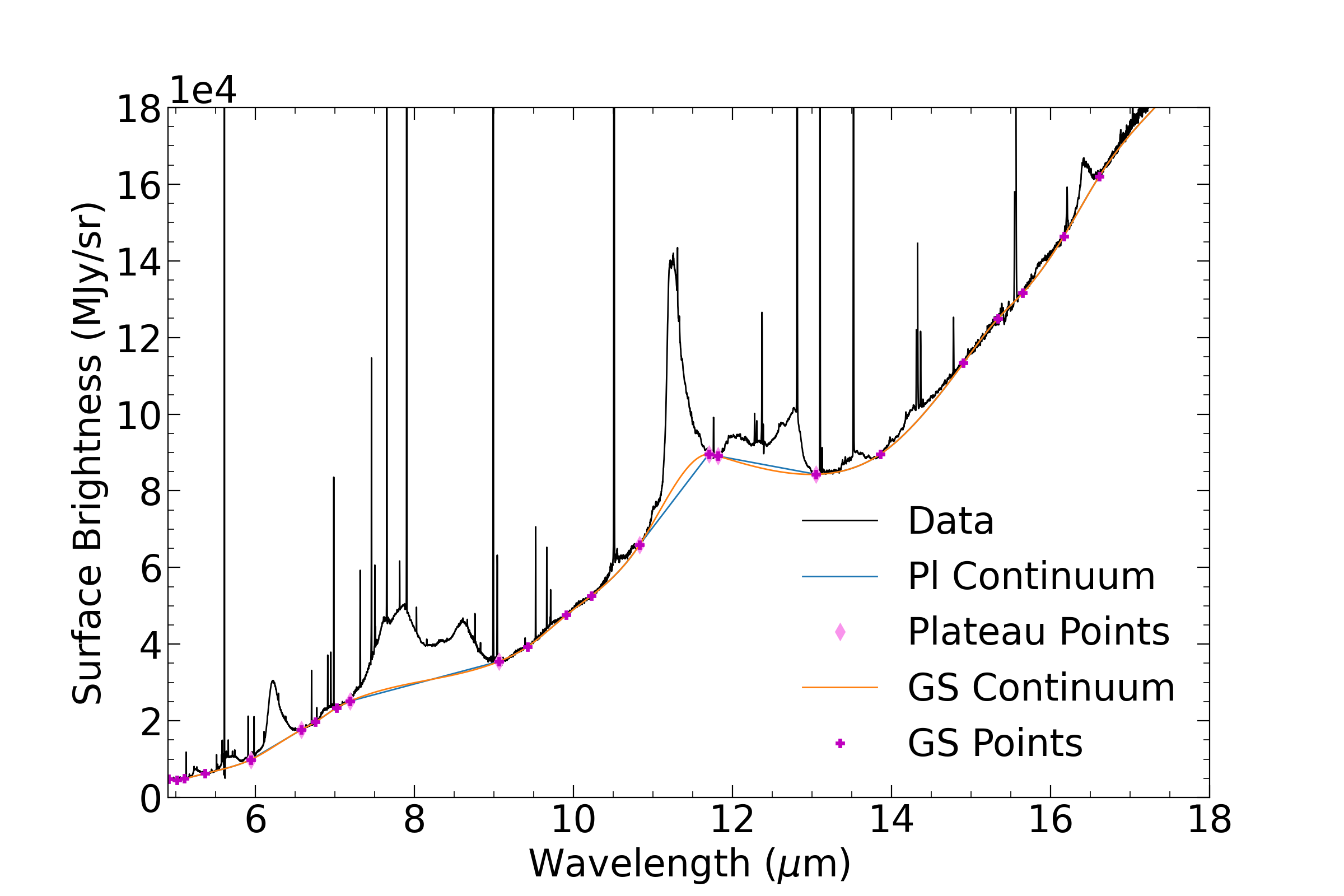}}
    \resizebox{.8\columnwidth}{!}{
    \includegraphics[width=\linewidth, clip, trim=1.5cm 0cm 2cm 1.5cm]{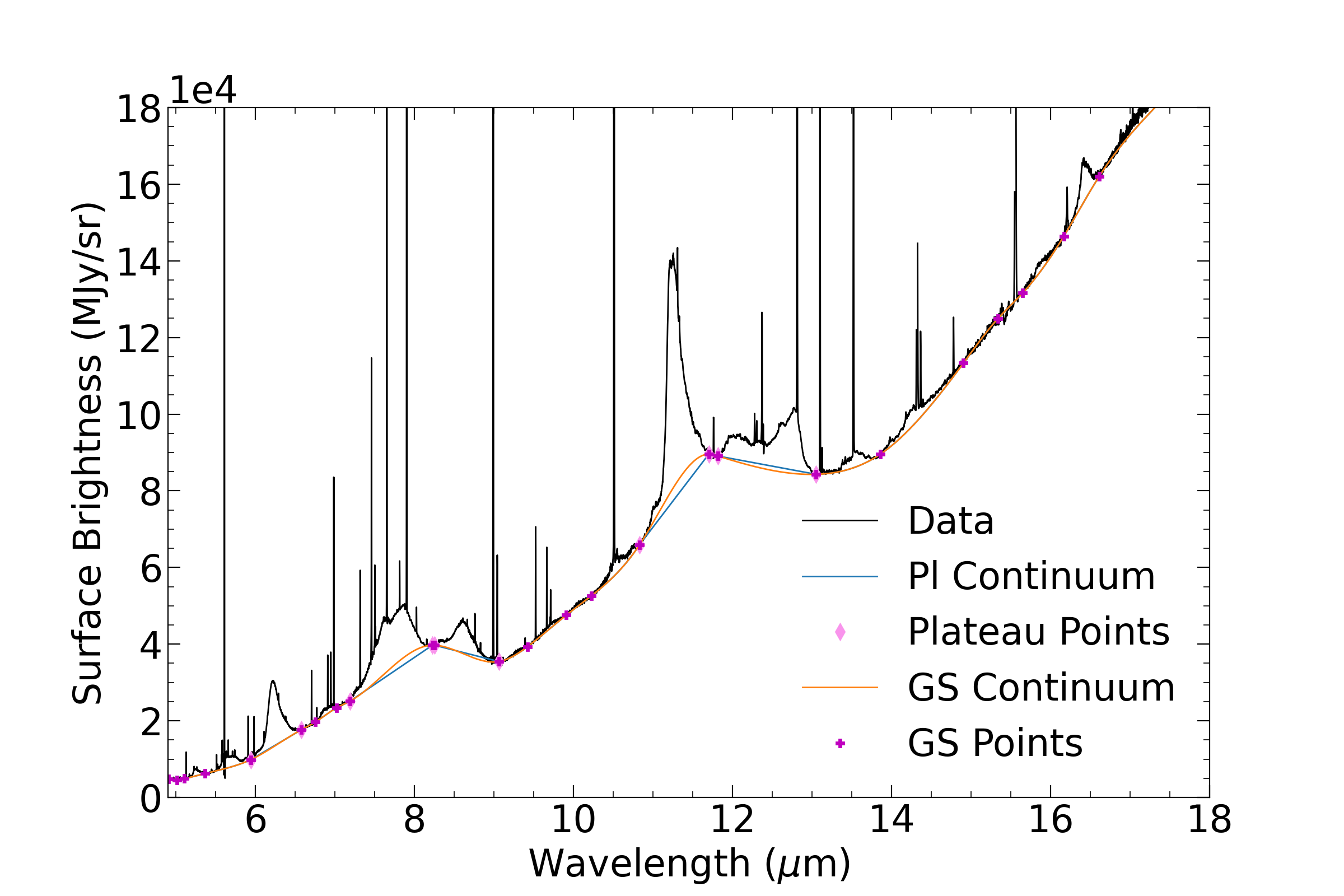}}
    \caption{\small Illustration of the global spline (GS, orange) and plateau (PL, green) continuum without (top) and with (bottom) an anchor point near 8.2~\mum for the MIRI-MRS integrated spectrum of NGC~7027.
    }
\label{fig:cont}
\end{figure}

In order to isolate the PAH emission, the rising dust continuum emission and the emission from other contributors must be removed from the spectra. To trace out the continuum emission, we place anchor points at wavelengths solely probing the dust continuum emission and on either side of each PAH band. 
We estimate the continuum emission for the MIRI-MRS spectral cube in two ways: one version has anchor points at 8.225 and 8.25~\mum and another without anchor points around 8.2~\mum  (Fig. \ref{fig:cont}). In addition, we fit both spectra with a plateau (PL) continuum, which is simply a linear fit between any two specified points and a spline curve (global spline, GS) elsewhere. The PL continuum is used to fit linearly underneath the main PAH features to avoid unusual effects from the spline curve which artificially augment or diminish the strength of the PAH emission. We thus use these PL continua for all plots and analysis. 
Following the original classification of \citet{Peeters:prof6:02}, we use the PL continuum with anchor points at $\sim$8.2~\mum for the profile classification of the 7.7 and 8.6~\mum PAH bands. For all other analyses, the PL continuum without the anchor points at $\sim$8.2~\mum is used.

\section{PAH Profile Classes of the MIRI-MRS Integrated Spectrum of NGC~7027}
\label{app:sec:MIRI_integrated_spectrum}

\begin{figure}[htbp]
    \centering
    \resizebox{.95\columnwidth}{!}{\includegraphics[width=\linewidth]{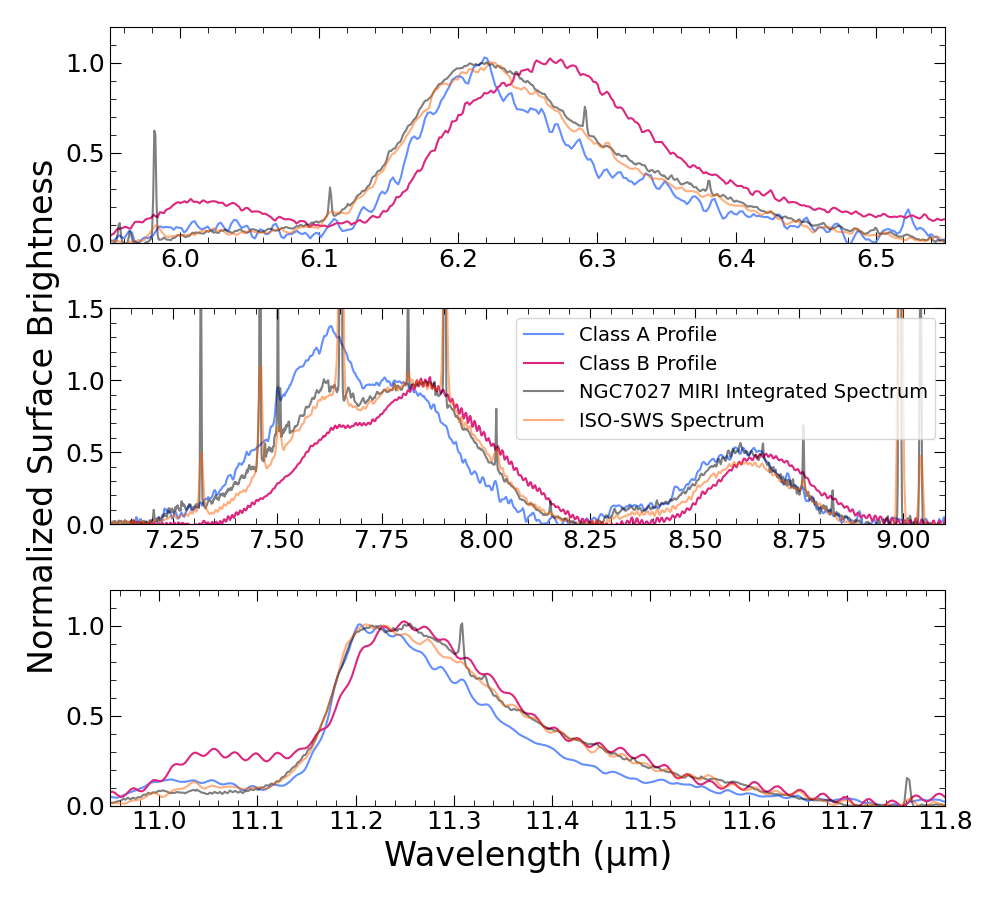}}
    \caption{\small 
    The MIRI-MRS integrated spectrum of NGC~7027 along with the PAH profile class \classA (blue) and \classB (red) for the 6.2, 7-9, and 11.2~\mum features 
    \citep[exemplified by IRAS~23133+6050 and HD~44179 respectively; taken from][]{Peeters:prof6:02, vanDiedenhoven:chvscc:04}.}
    \label{fig:miri_integrated_sp_classes}
\end{figure}

Fig.~\ref{fig:miri_integrated_sp_classes} shows the MIRI-MRS integrated spectrum of NGC~7027 along with archetypal spectra for PAH profile classes \classA and \classB for each of the 6.2, 7.7, 8.6, and 11.2~\mum features. The MIRI integrated spectrum shows a mix of PAH profile classes. The 6.2 and 8.6~\mum PAH profiles exhibit a class \classA profile whereas the 7.7 and 11.2~\mum features show a class \classAB profile. 

\section{Profile Variability Across NGC~7027}

\begin{table}[]
\caption{Overview of the PAH spectral classes in NGC~7027.}
   \centering
   \begin{tabular}{llll}
    \hline
    \multicolumn{1}{c}{Class} & \multicolumn{1}{c}{Position} & \multicolumn{1}{c}{FWHM} & \multicolumn{1}{c}{Region} \\
    & \multicolumn{1}{c}{(\mum)} & \multicolumn{1}{c}{(\mum)} &  \\
\hline\\[-7pt]
    \multicolumn{4}{c}{\textbf{6.2~\mum PAH}}\\    
    $\mathcal{A}$ & 6.19-6.23 & $\sim$0.14 & ISO-SWS\tablefootmark{b}, Outer NE \\ & & & Corner \\
    $\mathcal{AB}$ & 6.19-6.23 & $\sim$0.165 & MIRI-MRS\tablefootmark{a}, Inner Region \\
    $\mathcal{B}$ & >6.24 & $\sim$0.165 & E Ring, SW Ring \\[2pt]

   \multicolumn{4}{c}{\textbf{7.7~\mum PAH}}\\ [2pt]   
    $\mathcal{A}$ & $\sim$7.6 & -- & Outer NE Corner \\
    $\mathcal{AB}$ & $\sim$7.6 and $\sim$7.8 & -- & MIRI-MRS\tablefootmark{a}, Inner Region \\
    $\mathcal{B}$ & $\sim$7.8 & -- &  ISO-SWS\tablefootmark{b}, E Ring, \\ & & & SW Ring \\[2pt]

    \multicolumn{4}{c}{\textbf{8.6~\mum PAH}}\\[2pt]   
    $\mathcal{A}$ & 8.58-8.62 & -- & MIRI-MRS\tablefootmark{a}, Outer NE \\ & & & Corner, Inner Region \\
    $\mathcal{B}$ & >8.62 & -- & ISO-SWS\tablefootmark{b}, E Ring, \\ & & & SW Ring \\
[2pt]
   \multicolumn{4}{c}{\textbf{11.2~\mum PAH}}\\  [2pt]  
    $\mathcal{A}$ & 11.20-11.24 & $\sim$0.17 & Outer NE Corner \\
    $\mathcal{AB}$ & 11.20-11.24 & $\sim$0.21 & ISO-SWS\tablefootmark{b}, SW Ring\\
    $\mathcal{B}$ & $\sim$11.25 & $\sim$0.20 & MIRI-MRS\tablefootmark{a}, E Ring, \\ & & & Inner Region  \\

    \hline
    \end{tabular}
    \tablefoot{
    \tablefoottext{a}{MIRI-MRS spectrum integrated over channel 1 FOV (see Fig.~\ref{fig:fov}).} \tablefoottext{b}{FOV of 14" $\times$ 20".}}
    \label{tab:prof_summary}
\end{table}

Table~\ref{tab:prof_summary} demonstrates the rich variety of the PAH spectral characteristics of the ISO-SWS, MIRI-MRS, and spaxel apertures of NGC~7027 (Fig.~\ref{fig:fov}; Table~\ref{tab:aperture_info}). The outer NE corner is consistently characterized by class \classA profiles, while the E and SW ring are consistently characterized by class \classB profiles, with the exception of the 11.2~\mum spectrum in SW ring, which is class \classAB. The spectra of ISO-SWS, MIRI-MRS, and the inner region, exhibit variation in spectral classes across the distinct PAH bands.

\begin{figure}[htbp]
    \begin{center}
      \resizebox{\hsize}{!}{
    \includegraphics[ clip, trim=2.6cm 2cm .6cm 1cm]{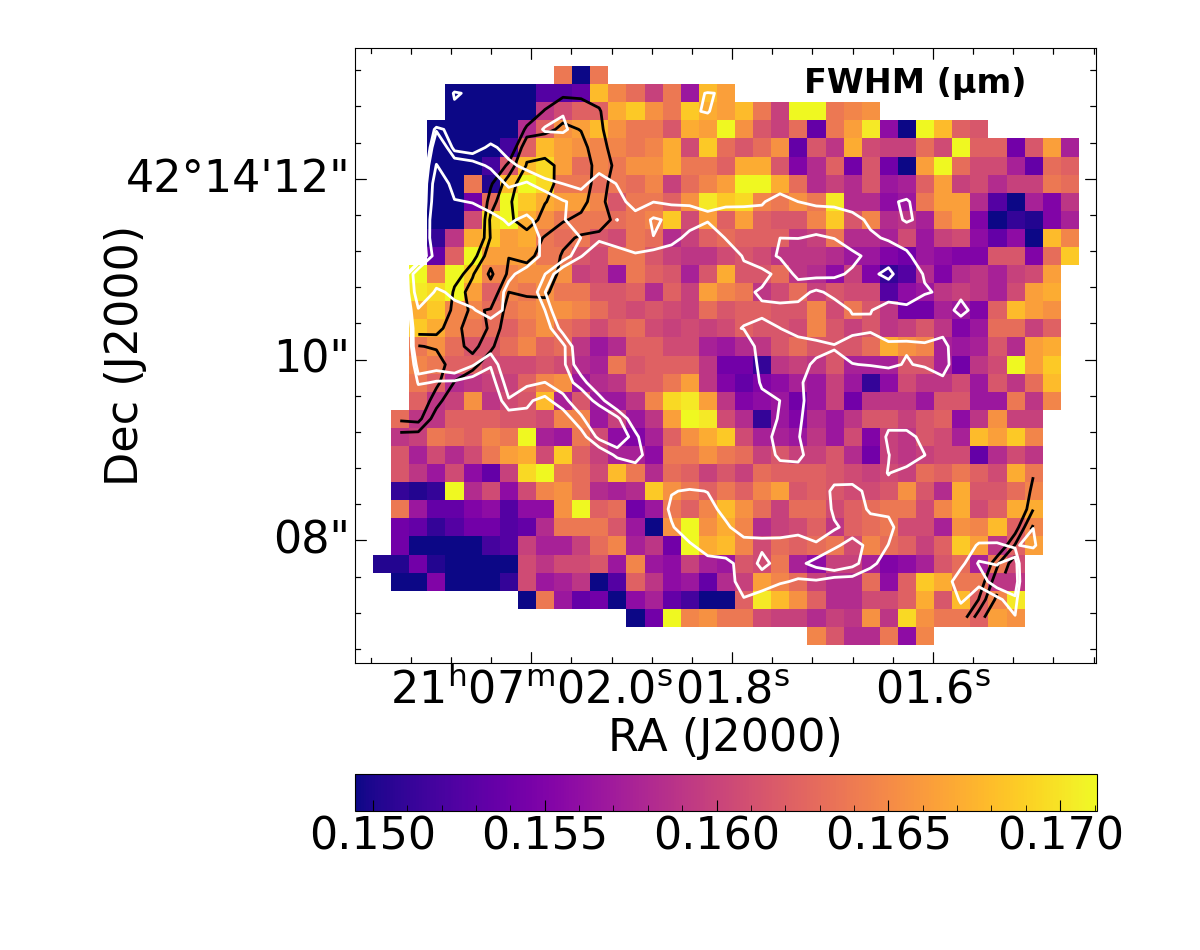}
\includegraphics[ clip, trim=2.6cm 2cm 1.2cm 1cm]{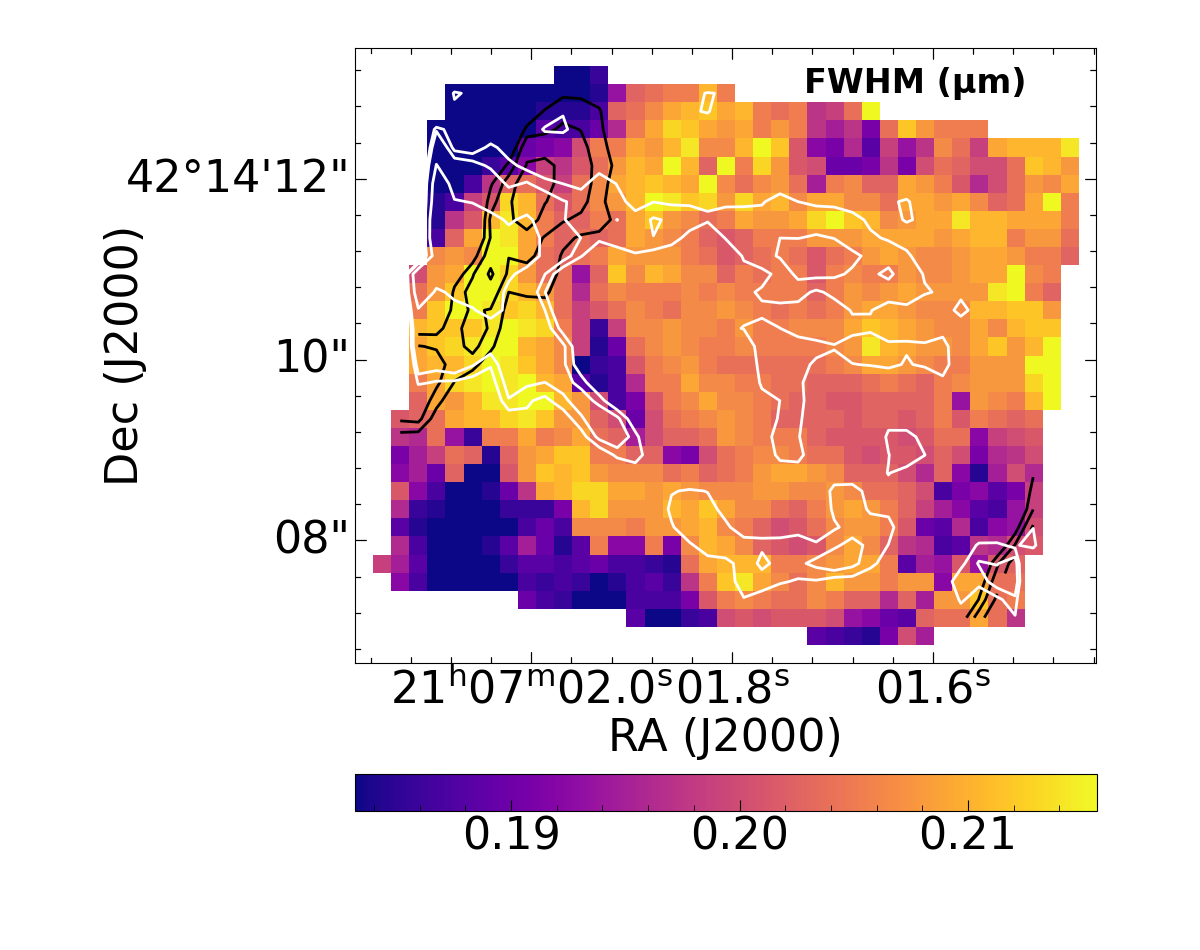}}

    \caption{\small Profile width variability of the 6.2 (left) and 11.2~\mum (right) features. Contours are as in Fig.~\ref{fig:profile-variability_maps}.}
    \label{fig:fwhm_maps}
    \end{center}
\end{figure}

The 6.2 and 11.2~\mum profile FWHMs across the nebula are measured. Fig.~\ref{fig:fwhm_maps} displays the changes in the profile widths across NGC~7027. Interestingly, the FWHM map of the 11.2~\mum feature roughly follows the \molh contours.

\section{Feature Decompositions}
\label{app:sec:decompositions}
Motivated by the decomposition performed on the 11.2~\mum band by \cite{Khan2025}, the 6.2~\mum band is decomposed using the same method. We use the narrowest profile in the MIRI-MRS cube at 6.205~\mum as the first component, and subtract this (scaled) component from the broadest profile in our mosaic to reveal component 2 which has a peak position of 6.26~\mum (Fig.~\ref{fig:62comps}). Subsequently, we fit the 6.2~\mum profile at each spaxel with a linear combination of these two components. This decomposition accurately models the 6.2~\mum profiles observed in the mosaic (Fig.~\ref{fig:62comps}). The integrated surface brightnesses of each component are measured and used for analysis.

\begin{figure}[htbp]
    \centering
     \includegraphics[width=\linewidth, clip, trim=0cm 0cm 3cm 2cm]{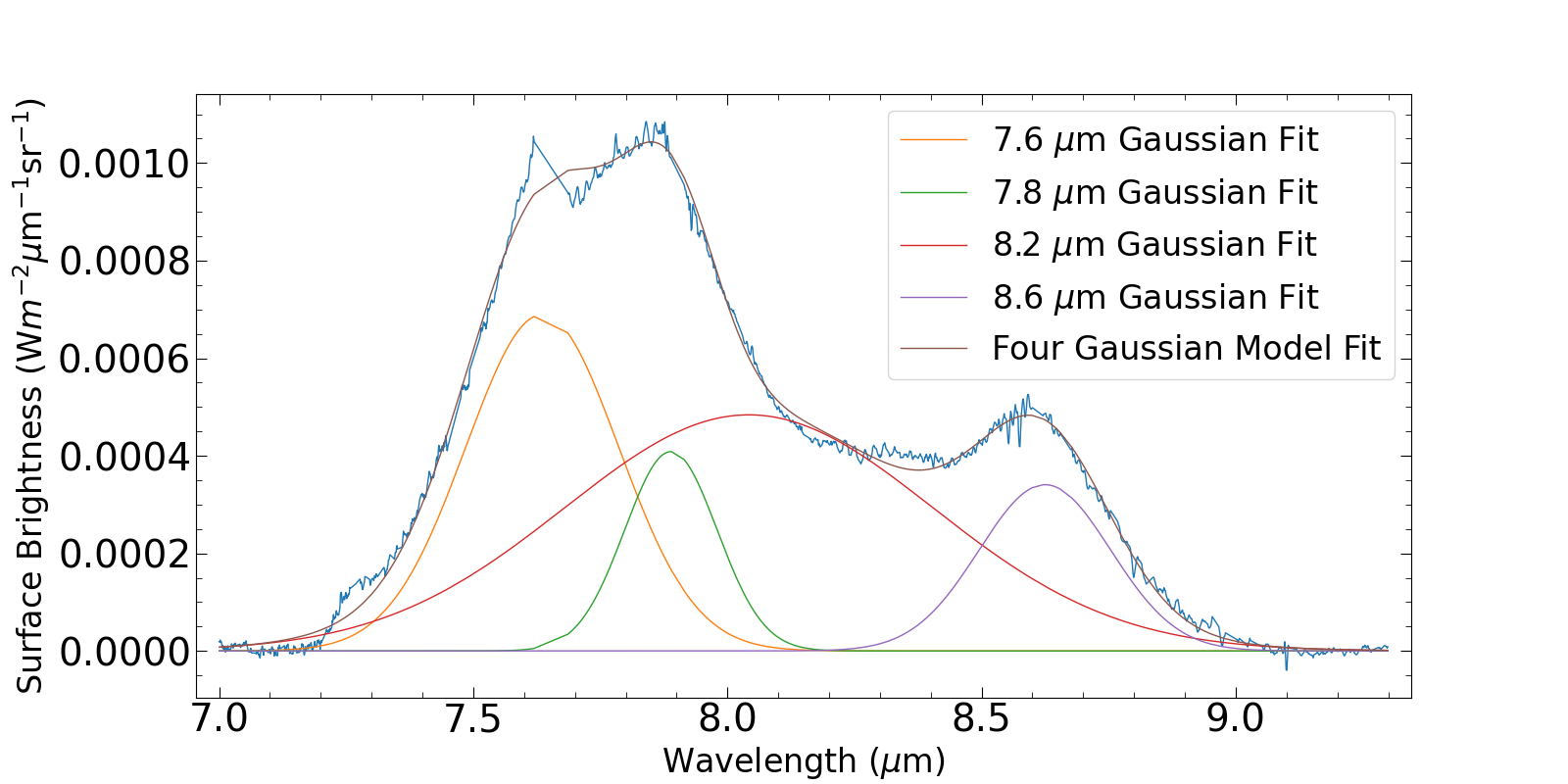}
    \caption{\small Four-Gaussian decomposition of the 7-9~\mum complex employing a broad G8.2 profile for an example spectrum. Components are G7.6, G7.8, G8.2, and G8.6~\mum.} 
    \label{fig:7-9GaussDecompositions}
\end{figure}

To reveal the contribution from each of the main PAH features within the 7-9~\mum complex across the nebula, we fit the 7-9~\mum complex using four Gaussian profiles to model the emission from the 7.6, 7.8, 8.2, and 8.6~\mum features following the method used in \citet{peeters2017} and \citet{Stock2017}. We follow the methodology employed by \citet{Stock2017} to determine the peak position and FWHM of these Gaussians. The fitting routine tends towards a broad G8.2 component, rather than a narrow G8.2 (as done by \citet{peeters2017} and \citet{Stock2017}). Subsequently, we fix the peak position (FWHM) to 7.636 (0.354), 7.889 (0.215), 8.042 (0.853), and 8.625 (0.305)~\mum, while we allow the amplitude to vary for each spectrum.  We obtain very good fits to the spectra using a broad G8.2~\mum component as shown in Fig.~\ref{fig:7-9GaussDecompositions}, where the 8.2~\mum component seems to represent both the 8.2~\mum feature employed by these authors and the underlying 7-9~\mum plateau upon which these four PAH bands sit.

Given the variation in the 8.6~\mum profiles, we perform a similar decomposition on this feature as done for the 6.2~\mum band to reveal the potential underlying components. Here, we use the most red-shifted profile in the MIRI-MRS cube at 8.65~\mum as the first component, and subtract this (scaled) component from the profile exhibiting the largest blue-shift to uncover component 2 with a peak position of 8.56~\mum (Fig.~\ref{fig:86comps}). We extract the 8.35~\mum feature which sits on top of the 8.56~\mum component by subtracting a linear continuum underneath it. Subsequently, we fit the 8.6~\mum profile at each spaxel with a linear combination of these two components and the 8.35~\mum feature. The observed 8.6~\mum profiles are well-modelled and the integrated surface brightnesses of each component are utilized for analysis.

Following the method in \cite{Khan2025}, we consider the narrowest 11.2 profile in our mosaic as the primary component, which peaks at 11.204~\mum, whereas in the OB the 11.207~\mum component peaks at 11.20~\mum. Second, we consider the broadest 11.2 profile in the spectral cube and subtract the scaled component 1 to obtain component 2, namely the 11.25~\mum component, which peaks at 11.29~\mum in NGC~7027 and in the OB \citep{Khan2025}. It is important to note that the primary component found here in NGC~7027 is broader than that of the OB, likely due to residual contamination from the secondary component, even in the narrowest profile. We attribute the slight differences in peak position to the decomposition method. Since the widest profile in NGC~7027 is broader than the widest profile in the OB, the 11.25~\mum component here is also slightly broader than in the OB.

Using a linear combination of these two weighted components, we successfully decompose the 11.2~\mum profile of each spaxel in the mosaic (see Fig.~\ref{fig:112comps}) and subsequently integrate the components to measure their integrated surface brightness spatially. 

\section{Feature Correlations}

\begin{figure}[htbp]
    \centering
    \includegraphics[width=\linewidth]{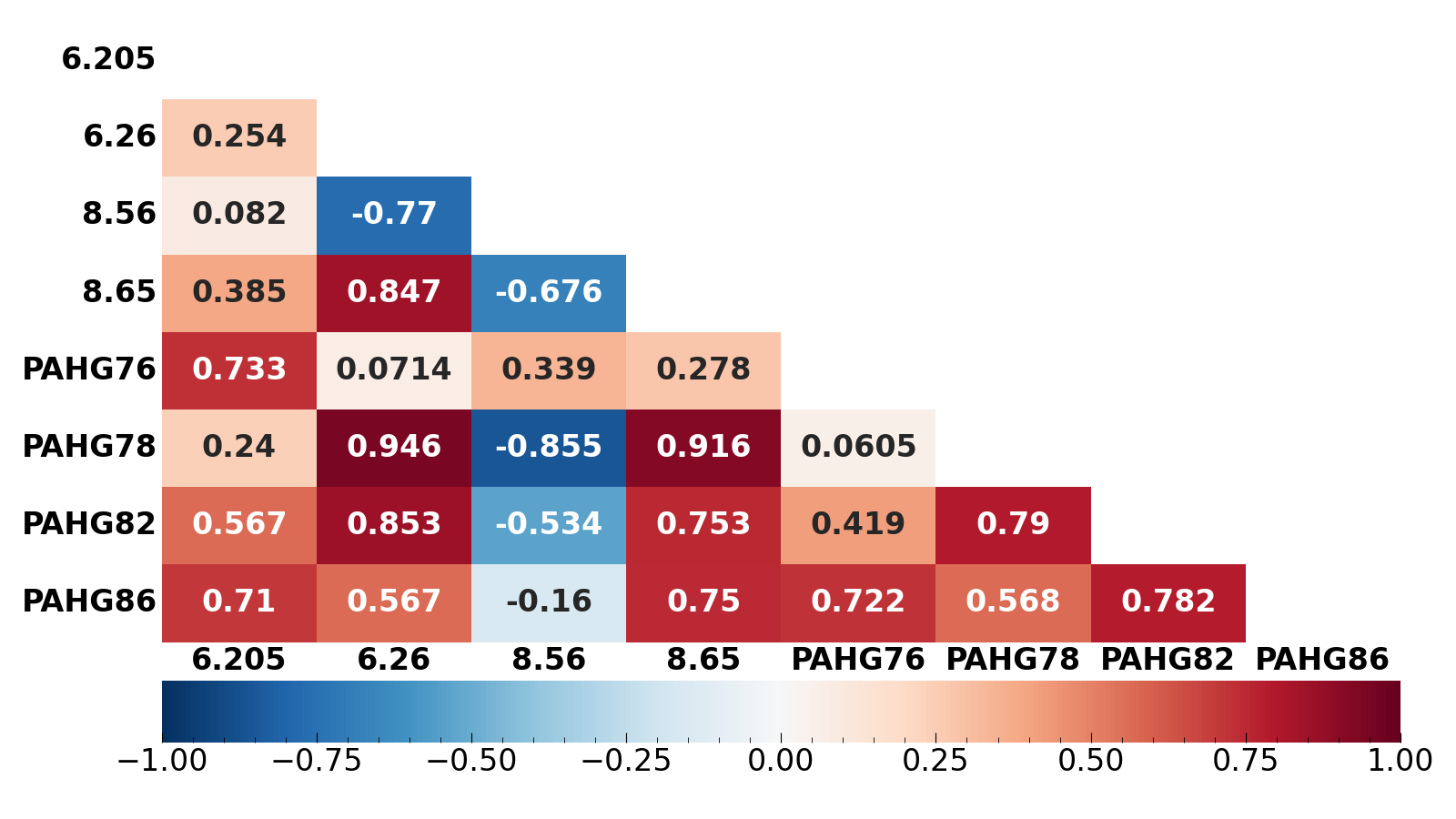}
    \caption{\small Heat-map demonstrating the strength of correlations between the 6-9~\mum components when normalized to the 11.207~\mum PAH component, where the intersection between two components displays their correlation coefficient. Redder boxes indicate a strong correlation, bluer boxes indicate a strong anti-correlation, and light boxes are present when there is little-to-no correlation.}
    \label{fig:heatmap}
\end{figure}

The heat-map highlighting the Pearson correlation coefficients between the 6-9~\mum components is shown in Fig.~\ref{fig:heatmap}. All features are normalized over the 11.207~\mum PAH component to remove any correlative behaviour that is simply due to the contribution of the total PAH emission since we aim to uncover the underlying relationships between PAH components.

\begin{figure}[htbp]
    \centering
    \includegraphics[width=\linewidth]{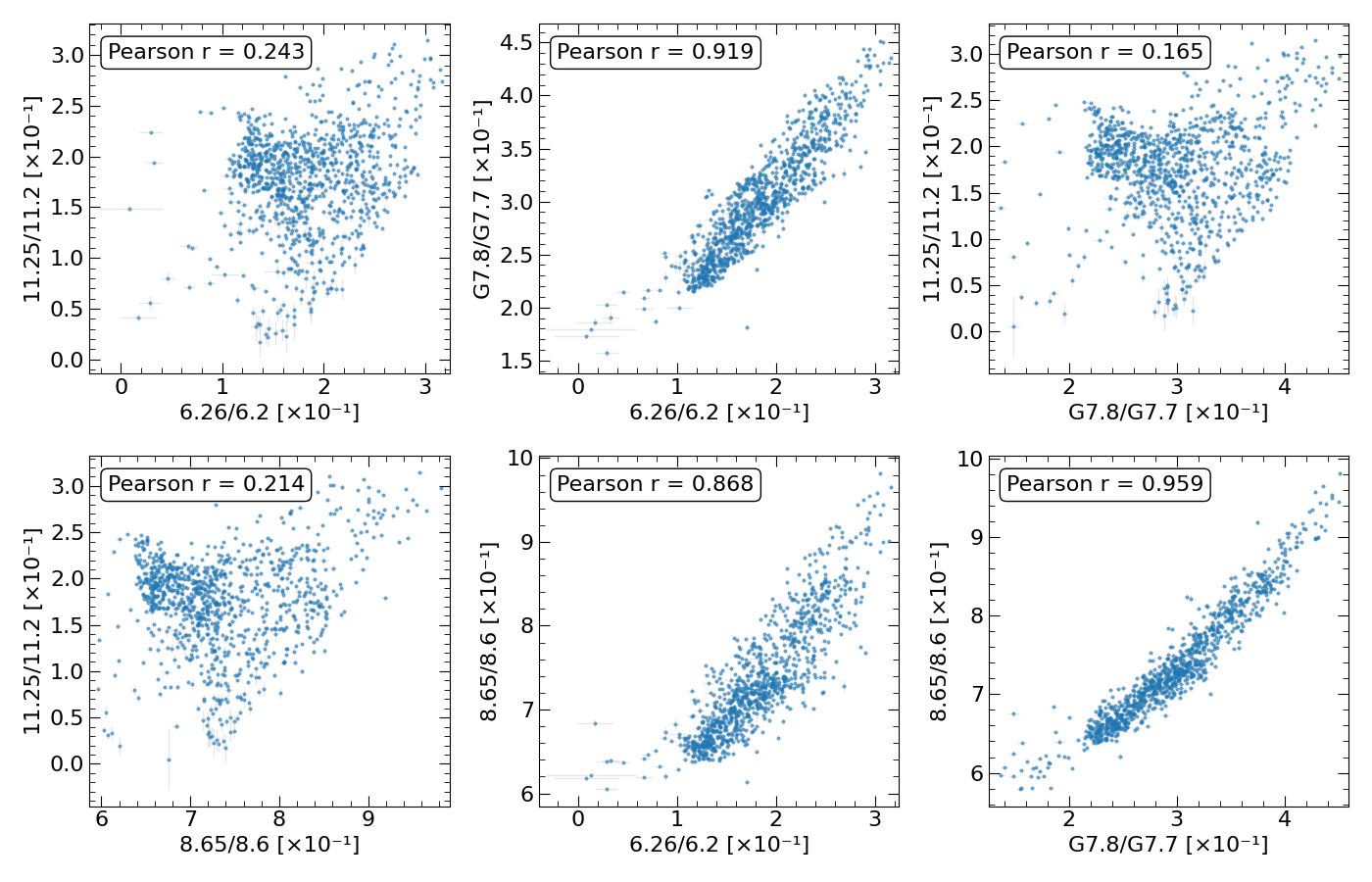}
    \caption{\small Correlation plots between the 6.26, 7.8, 8.65, and 11.25~\mum components normalized over their feature totals. The Pearson correlation coefficient is shown in the top left corner of the plots. Error bars are shown in light grey.}
    \label{fig:comp_correlations_norm_over_total}
\end{figure}

Figure~\ref{fig:comp_correlations_norm_over_total} shows no significant correlation between 11.25 and the red components of the 6.2, 7.7, and 8.6~\mum features. On the other hand, the 6.26, G7.8, and 8.65~\mum components demonstrate tight correlations with each other. Therefore, these plots serve as another indication that the 11.25~\mum component exhibits unique behaviour from the PAHs, while the 6.26, 7.8, and 8.65~\mum components belong to the same PAH subpopulation.

\section{PAH Band Ratio Maps}

The ratio of the integrated surface brightness of PAH bands are calculated to reveal the relative strength of the features spatially. In particular, we calculate the 6.2/11.207 PAH band ratio since this ratio this is a known tracer of PAH charge. We use the 11.207~\mum component since it contains only the PAH contribution to the 11.2~\mum, without significant contamination from the 11.25~\mum VSG/cluster component. The map reveals that, in general, the ionic PAHs dominate in the ring more than anywhere else. The ratio map of the G7.8/G7.6 PAH components is calculated and reveals a roughly co-spatial relationship with the 6.2/11.207 map.

\begin{figure}[htbp]
    \centering
    \includegraphics[width=\linewidth, clip, trim=0cm 2.5cm 0cm 2cm]{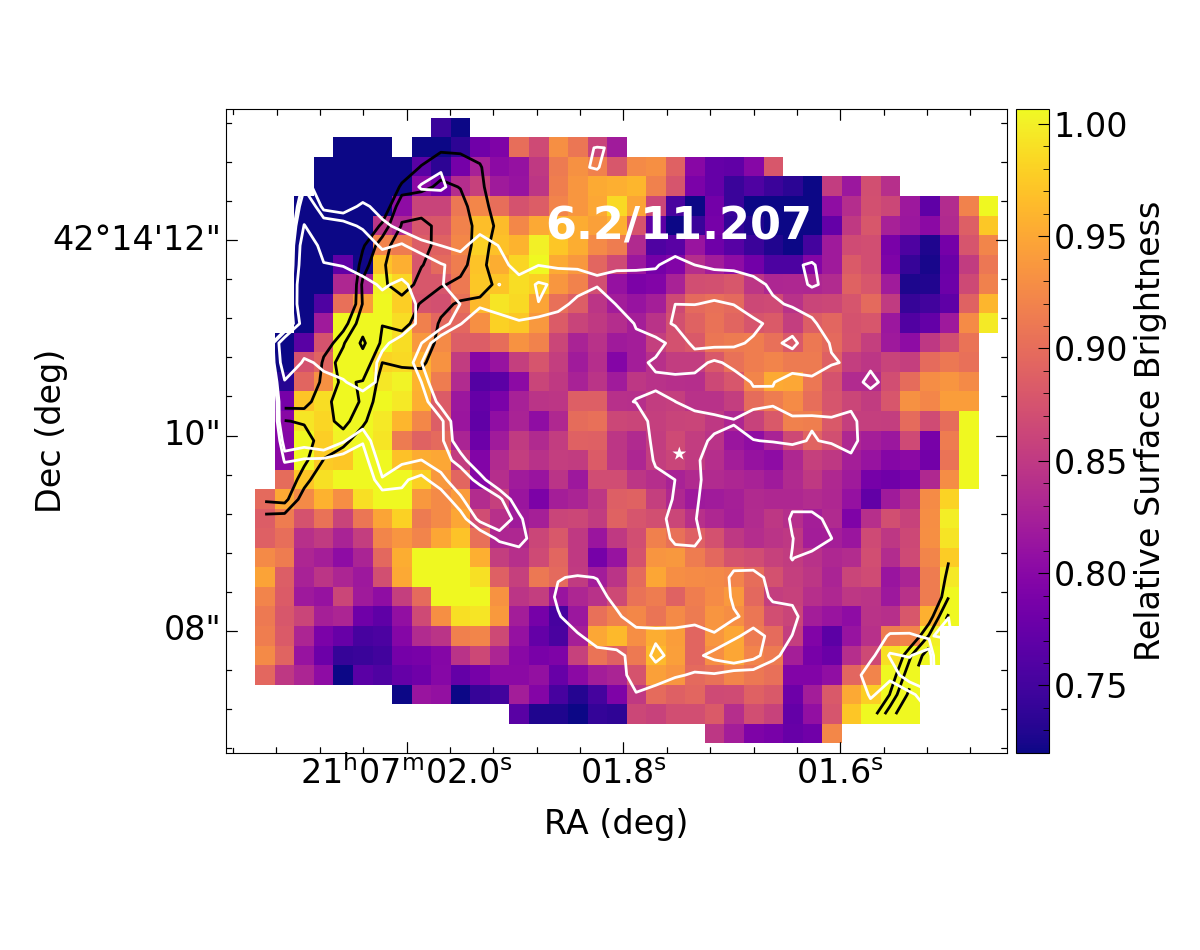}
    \caption{\small Spatial map of the 6.2~\mum feature over the 11.207~\mum component ratio. Contours are as in Fig.~\ref{fig:profile-variability_maps}.}
    \label{fig:62over11207_ratio_map}
\end{figure}

\begin{figure}[htbp]
    \centering
    \includegraphics[width=\linewidth, clip, trim=0cm 2.5cm 0cm 2cm]{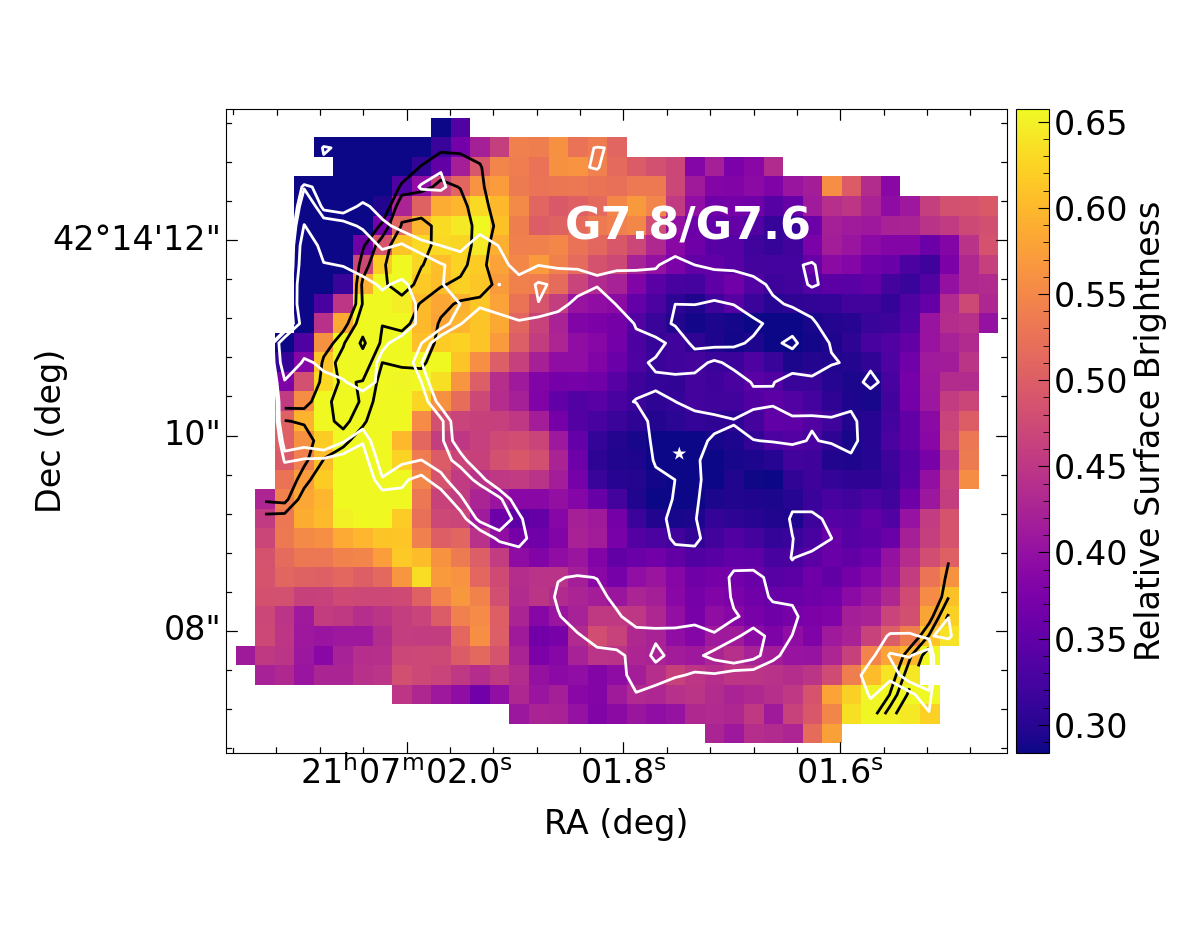}
    \caption{\small Spatial map of the G7.8~\mum component over the G7.6~\mum component ratio. Contours are as in Fig.~\ref{fig:profile-variability_maps}.}
    \label{fig:78over76_ratio_map}
\end{figure}

\section{Theoretical 6.2 and 7.7~\mum Band Positions}
\label{app:sec:band_positions}

For a set of small PAHs, the mid‑infrared (5-18~\mum) region has been studied in a cold molecular beam and modeled using anharmonic theory \citep{Lemmens2019}. Comparison between the FELIX free‑electron laser experimental spectra and second‑order vibrational perturbation theory calculations shows agreement in peak positions within $\sim$0.5\%, on average. Given that the FELIX bandwidth is $\sim$1\% of the photon energy, the theoretical values may in some cases be more accurate than the experimental determinations.

\end{appendix}

\end{document}